# Embedded, micro-interdigitated flow fields in high areal-loading intercalation electrodes towards seawater desalination and beyond

Vu Q. Do,[a] Erik R. Reale,[a] Irwin Loud IV,[a] Paul G. Rozzi,[a] Haosen Tan,[b]

David A. Willis,[b] and Kyle C. Smith[a,c,d,e,*]

a. Department of Mechanical Science and Engineering,
University of Illinois at Urbana-Champaign, Urbana, IL, USA

b. Department of Mechanical Engineering,
Southern Methodist University, Dallas, TX, 75275

c. Department of Materials Science and Engineering,
University of Illinois at Urbana-Champaign, Urbana, IL, USA

d. Computational Science and Engineering Program,
University of Illinois at Urbana-Champaign, Urbana, IL, USA

e. Beckman Institute for Advanced Science and Technology,
University of Illinois at Urbana-Champaign, Urbana, IL, USA

*corresponding author: kcsmith@illinois.edu

**Abstract**

Faradaic deionization (FDI) is a promising technology for energy-efficient water desalination using porous electrodes containing redox-active materials. Herein, we demonstrate for the first time the capability of a symmetric FDI flow cell to produce freshwater (<17.1mM NaCl) from concentrated brackish water (118mM), to produce effluent near freshwater salinity (19.1 mM) from influent with seawater-level salinity (496 mM), and to reduce the salinity of hypersaline brine from 781 mM to 227 mM. These remarkable salt-removal levels were enabled by using flow-through electrodes with high areal-loading of nickel hexacyanoferrate (NiHCF) Prussian Blue analogue intercalation material. The pumping energy consumption due to flow-through electrodes was mitigated by embedding an interdigitated array of <100 μm wide channels in the electrodes using laser micromachining. The micron-scale dimensions of the resulting embedded, micro-interdigitated flow fields (eμ-IDFFs) facilitate flow-through electrodes with high apparent permeability while minimizing active-material loss. Our modeling shows that these eμ-IDFFs are more suitable for our intercalation electrodes because they have >100X lower permeability compared to common redox-flow battery electrodes, for which millimetric flow-channel widths were used exclusively in the past. Total desalination thermodynamic energy efficiency (TEE) was improved by more than ten-fold relative to unpatterned electrodes: 40.0% TEE for brackish water, 11.7% TEE for hypersaline brine, and 7.4% TEE for seawater-salinity feeds. Water transport between diluate and brine streams and charge efficiency losses resulting from (electro)chemical effects are implicated as limiting energy efficiency and water recovery, motivating their investigation for enhancing future FDI performance.

**Broader Context**

During the past decade, the field of electrochemical separations has experienced a surge of research activity stimulated by the introduction of redox-active electrode materials, as motivated by their potential to increase salt removal, energy efficiency, and selectivity toward applications that support global sustainability, including desalination, $CO_2$ capture, environmental remediation, and nutrient/resource recovery. However, past experimental desalination studies with substantial salt removal have been limited to brackish water feeds when flow cells were used, despite our early modeling [Smith and Dmello, *J. Electrochem. Soc.*, **163**, A530 (2016)] that predicted seawater desalination using cells comprised of symmetric cation intercalation electrodes, separated by cation-blocking membranes. This work is the first to demonstrate experimental salt removal approaching seawater salinity by using high areal-loading flow-through intercalation electrodes that are engraved with interdigitated microchannel networks to minimize electrical and pumping energy consumption. This new paradigm for the structuring of porous electrodes motivates the further development of embedded microfluidic networks and their use in various electrochemical processes with liquid or gaseous feeds, including flow batteries for energy storage, fuel cells for energy conversion, and electrochemical $CO_2$ conversion.

**Introduction**

With the global human population growing at an ever increasing rate and demanding more water for domestic, agricultural, and industrial purposes, desalination of brackish water and seawater is considered an alternative to freshwater resources and is expected to produce more than 200 million m$^3$ per day in 2030.[1] Reverse osmosis (RO) presently contributes nearly 70% of global desalination capacity, followed by thermal-based processes with ~25%.[2,3] However, thermal distillation is energy intensive[4] due to the heat of vaporization of water. While RO is energy efficient,[5] its water recovery is mechanically limited by the pressure necessary to overcome osmosis into highly concentrated saltwater.[6] Further, both such processes result in the indiscriminate removal of salt, thus requiring remineralization.[7,8] In addition, RO plants require large capital investment, substantial maintenance, and present environmental impacts caused by pretreatment chemicals and membrane cleaning.[9]

Desalination that uses electric fields to transport ions, rather than using pressure to transport water molecules as in RO, promises reduced brine volume and environmental impact. Hybrid RO systems that employ electrodialysis to increase water recovery have been analyzed[10,11] and demonstrated,[12,13] but electroosmosis through electrodialysis stacks was shown to limit the degree of brine concentration.[10] Further, the use of gas-evolution electrodes in electrodialysis stacks with only a few membrane pairs results in large energy consumption due to the large difference in electrode potentials (~1 V when using $H_2$ and $O_2$ evolution). Alternatively, capacitive deionization (CDI) using ion adsorption in electric double layers (EDLs) has achieved better energy efficiency than RO at brackish salt concentrations (<10 g L$^{-1}$ or 171mM NaCl)[14] with much research focused on developing high-capacity, high-rate electrode materials.[15,16] Even with advanced materials, salt adsorption capacity (SAC) using EDLs rarely exceeds 15 mg per gram of active material,[17] and thermodynamic energy efficiency (TEE) remains below 10%.[18,19] This is after modifications to CDI process design that have yielded greater improvements in SAC and TEE, demonstrating the importance of system optimization.[18] For example, using ion-exchange membranes in CDI can improve TEE by one order of magnitude,[19] and flow-electrode CDI can improve SAC by >50%.[20,21] Despite these advances, CDI using EDL charge storage has been relegated to brackish water desalination, mainly due to its low capacity.

In contrast, redox-active intercalation materials used in Faradaic deionization (FDI) promise seawater desalination on the basis of experiments with non-flowing cells[22,23] and earlier modeling with flowing cells.[24,25] This can be attributed to their high ion-storage concentrations (>4 mol $L^{-1}$ in Ref. 26) and large SACs (as large as ~100 mg $g^{-1}$ in Ref. 27). Despite this, the desalination of seawater salinity in a practical FDI flow cell using intercalation electrodes remains to be demonstrated experimentally. Past FDI studies[23,26,28–33] have used porous electrodes incorporating Prussian blue analogue (PBA) intercalation material at PBA mass loadings under 10 mg $cm^{-2}$ to remove less than 20% of the salt from feeds with seawater salinity. However, the salt removal from these studies shows a positive correlation between electrode capacity and salt removal (Fig. 1a). This observation suggests scaling up charge capacity as a means to approach seawater desalination, as we demonstrate subsequently by upsizing electrodes.

Aspects of device architecture other than capacity have significant impacts on SAC, salt adsorption rate (SAR), and TEE.[19,34] Among existing flow-cell configurations, symmetric FDI cells[24,25,35] have outperformed hybrid CDI, dual-ion desalination, and the desalination battery.[19,34] In addition, flow-through electrodes[25,26,28,30,36] have been shown to outperform flow-by electrodes.[25,30,36] The enhanced performance of flow-through electrodes is a result of the intimate contact of flowing salt solution with intercalation material that enables salt-depleted solution within electrodes to be evacuated and replaced efficiently with new salt-rich solution (see SI and Fig. S1). Therefore, flow-through configurations remove salt faster while yielding higher utilization of active material capacity. However, the fine porosity of electrodes containing intercalation nanoparticles (e.g., PBAs) produces large pumping losses in flow-through configuration because (1) pumping energy $E_{pump}$ scales with the square of pressure drop $\Delta p$ for a desired volumetric flow rate $Q$ under creeping-flow conditions ($E_{pump} \propto Q\Delta p \propto \Delta p^2$)[28] and (2) $\Delta p$ scales inversely with electrode hydraulic permeability $k_h$ that scales with the square of pore size. While past FDI studies have neglected pumping losses,[37] we have shown recently that pumping losses can exceed electrical energy input by 50-fold in flow-through FDI.[26] Large-area electrodes also result in increased pumping pressure due to increased flow path length, further suppressing desalination energy efficiency and further necessitating a means of its mitigation to make flow-through FDI viable.

In this work, we report symmetric FDI with flow-through nickel hexacyanoferrate (NiHCF) electrodes approaching the removal of seawater salinity levels, while achieving pumping losses at parity with or lower than electrical losses for the first time (Fig. 1c). These salt removal levels were achieved with high areal-loading electrodes having NiHCF loadings greater than 19 mg cm$^{-2}$, together with an automated valve-switching system to recirculate feed water so as to minimize state-of-charge gradients caused by streamwise polarization induced by salt concentration gradients.[26,36] Pumping losses were reduced by more than ten-fold by increasing hydraulic permeation through intercalation electrodes by embedding interdigitated microchannels within them via laser micromachining, resulting in novel embedded, micro-interdigitated flow fields (eµ-IDFFs). Past CDI studies using through-plane flow-through electrodes have avoided excessive pumping energy by using electrode materials with large macropores[38] and laser-perforated macrochannels.[39] However, such approaches are incompatible with FDI where an impermeable membrane is used to separate electrodes,[26,28,30] thus requiring in-plane fluid distribution.

For in-plane distribution of fluid through electrodes in redox-flow batteries[40,41] (RFBs) and fuel cells, various flow fields including interdigitated ones (IDFFs) have been used.[42] With one exception,[40] the past use of in-plane flow fields has embedded them in ~1 cm thick bipolar plates that compress monolithic carbon electrodes, whereas we presently embed them within intercalation electrodes that are cast on a ~100 µm graphite foil, thus yielding a low-profile design that minimizes the cost and mass of inactive cell components. Further, the apparent hydraulic permeability $k_h^{app}$ produced by embedding IDFF channels of width $w$ scales as $k_h^{app} \sim w^{-2}$ (see SI) when fixing the ratio of channel width to inter-channel spacing $s$, motivating IDFF miniaturization presently. Despite this, Fig. 1b shows that channel widths and inter-channel spacings of past IDFFs and other flow fields used for RFBs are ten-fold larger than the present eµ-IDFFs that we introduce and demonstrate for the first time. The eµ-IDFFs introduced here are shown to produce transverse flow between interdigitated microchannels less than 100 µm wide with as much as 100-fold enhanced permeability when embedded in intercalation electrodes, despite producing parallel flow and less than four-fold enhanced permeability when embedded in carbon-felt electrodes used in RFBs.

In what follows we first introduce rational design criteria for eµ-IDFFs using physics-based modeling of flow-through porous electrodes patterned with macro- and micro-porosity, identifying a key dimensionless parameter that justifies the use of microchannels despite the exclusive use of millimetric channels in earlier IDFFs for RFBs. Subsequently, the results of eµ-IDFF fabrication by laser micromachining are presented, where the imbibing of electrodes with water before laser micromachining is shown to minimize heat affected zones and improve channel quality. Next, we report results of desalination experiments using a symmetric FDI cell using eµ-IDFFs with feed-water recirculation. The pair of electrodes having the highest loading achieved 93.5%, 96% and 70.9% salt removal from brackish water (118 mM NaCl), seawater-salinity (496 mM), and hypersaline (781 mM) feeds, respectively. Because pumping energy was reduced significantly using eµ-IDFFs, TEE approached 40% for brackish water desalination and ~10% for seawater and hypersaline brine desalination. Desalination experiments using unpatterned electrodes were also conducted, showing 23% lower specific capacity compared to that of patterned electrodes, indicating that eµ-IDFFs help to minimize dead volumes within dense electrodes. Water recovery in excess of 50% is reported with water transport and charge efficiency loss mechanisms identified as future means of enhancement of desalination performance.

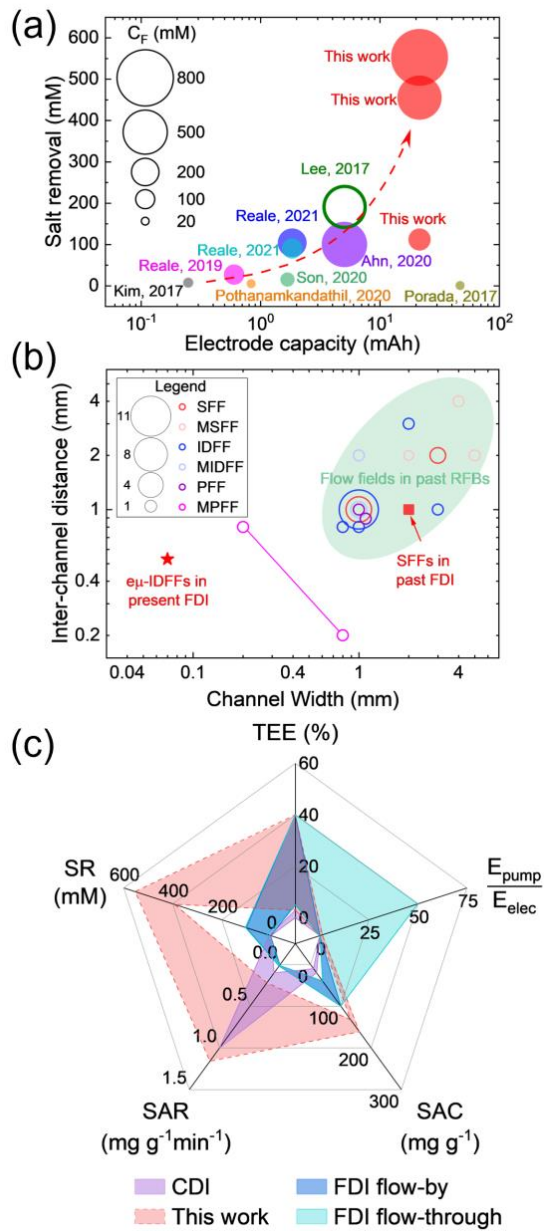

**Figure 1**. (a) Salt removal achieved presently compared to previous work using flow cells with NiHCF electrodes (Table S1). Area of a bubble is proportional to the feed concentration $C_F$ used in the work it represented. (b) Inter-channel distance and channel width for interdigitated (IDFFs), serpentine (SFFs), and parallel (PFFs) flow fields and modified versions thereof (MFFs) used presently and used previously in RFBs[43,44,53–62,45,63–67,46–52] and FDI[68] (Table S2). The area of a given bubble is proportional to the number of flow fields using such dimensions. The dumbbell symbol is for a corrugated MPFF having channel sections of alternating width and spacing.[69] (c) Radar plot comparing desalination metrics of the present work with those of past FDI and CDI work (Table S3).[17,19,23,26,28–33]

**Results and Discussion**

*Design and Fabrication of eµ-IDFFs*

We fabricated three pairs of electrodes with different nickel hexacyanoferrate (NiHCF) loading levels to embed them with eµ-IDFFs and to subsequently characterize their desalination performance in a symmetric FDI flow cell. After drying, all electrodes were found to be crack free (Fig. 2b) with thicknesses between 400 and 450 µm, and each was subsequently calendered down to ~200 µm (excluding the graphite foil which is ~100 µm thick). Next, laser micromachining was used to engrave eµ-IDFFs into these electrodes using the design shown in Fig. 2d. Properties of the six electrodes from these three pairs are shown in Table 1, where L, M, and H denote electrode pairs with lowest (~15 mg cm$^{-2}$), medium (~19 mg cm$^{-2}$), and highest (~21 mg cm$^{-2}$) NiHCF loading.

**Table 1**. Properties of electrodes after calendaring. The apparent permeabilities $k_h^{app}$ were measured after electrodes were engraved with eµ-IDFFs.

| Pair ID | Electrode ID | NiHCF loading (mg cm$^{-2}$) | Porosity, $\varepsilon$ (%) (uncalendared) | Porosity, $\varepsilon$ (%) (calendared) | Permeability, $k_h^{pe}$ (µm$^2$) | Apparent permeability, $k_h^{app}$ (µm$^2$) |
|---|---|---|---|---|---|---|
| L | 1 | 15.3 | 64.3 | 46.5 | 0.69 | 121.46 |
| | 2 | 15.3 | 64.3 | 46.5 | 1.06 | 52.99 |
| M | 3 | 19 | 62.0 | 42.2 | 0.63 | 60.98 |
| | 4 | 18.6 | 60.0 | 43.5 | 0.236 | 36.4 |
| H | 5 | 21.5 | 57.0 | 39.8 | 0.58 | 31.04 |
| | 6 | 21.3 | 57.4 | 40.4 | 0.48 | 31 |

The eµ-IDFFs that we have embedded in intercalation electrodes increase apparent hydraulic permeability by 30- to 120-fold (Table 1) by reducing the flow path length through electrodes by ~100-fold from 4.5 cm down to 500 µm (Fig. 2a). As we subsequently show, the width of microchannels must be sufficiently large to achieve uniform flow through porous-electrode regions while also being sufficiently small to minimize material loss due to the embedding of eµ-IDFFs. The former attribute is critical to ensure that salt-rich solution is supplied everywhere within the electrode to maximize intercalation-material utilization, whereas the latter is essential to maintain high charge capacity post-engraving.

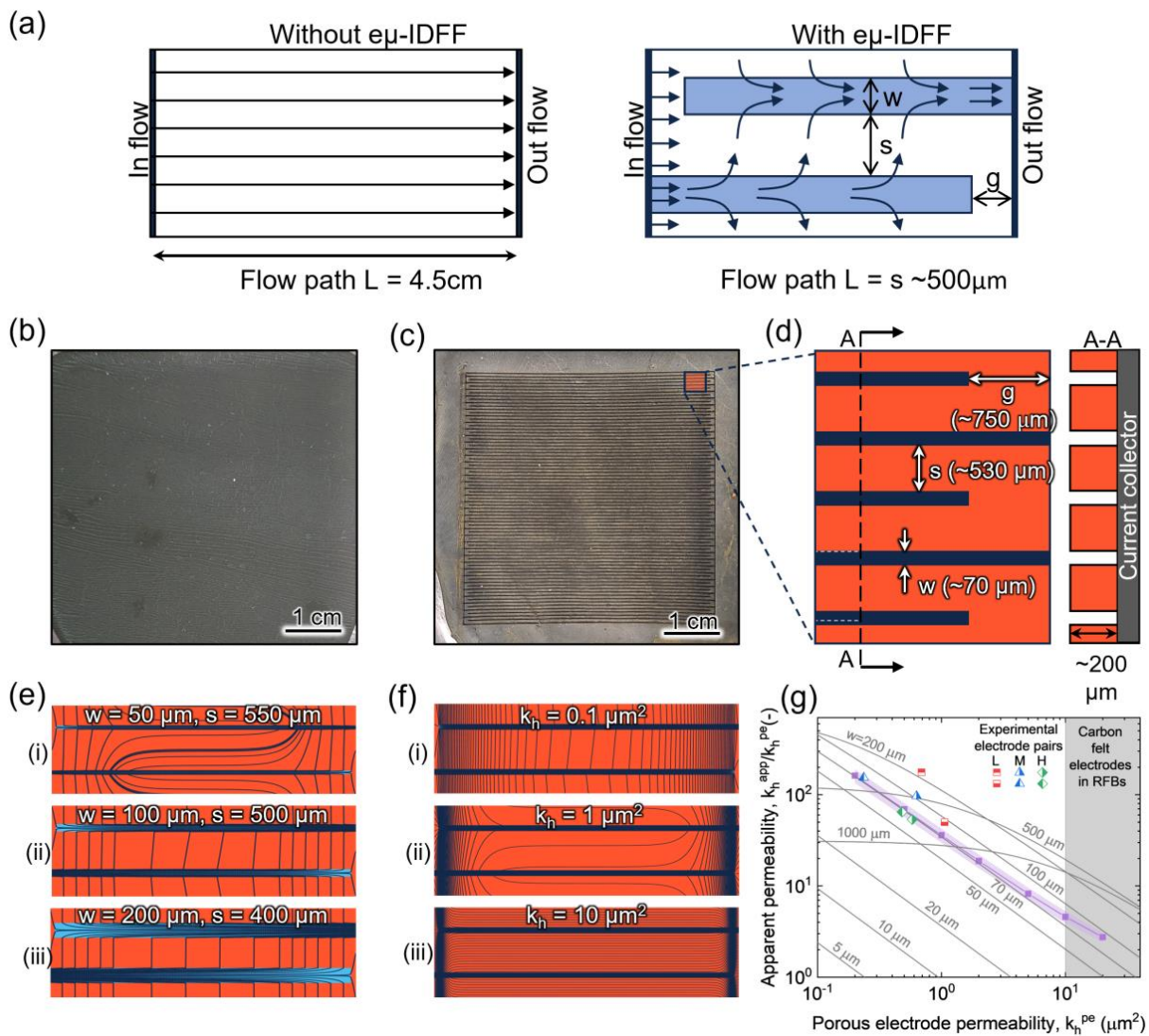

**Figure 2**. (a) Schematics comparing the flow path of electrodes with and without eµ-IDFFs. Photos of electrodes (b) after drying and (c) after being calendared and engraved with an eµ-IDFF. (d) Dimensions of the chosen eµ-IDFF design with a cross-section shown through the A-A plane. (e) Calculated streamlines for three cases with different channel width $w$ and inter-channel distance $s$. The gap $g$ and electrode permeability $k_h^{pe}$ were kept constant at 750 µm and 0.28 µm². (f) Calculated streamlines for three cases with the same channel dimensions shown in (d) but with electrode permeability varying. (g) Predicted and experimental ratios between the apparent

permeability of each patterned electrode $k_h^{app}$ to porous-electrode permeability $k_h^{pe}$. Purple symbols indicate data obtained from the numerical model using the channel dimensions shown in (d). Gray lines indicate data obtained using the quasi-1D analytical model for various channel widths with a fixed ratio of channel width to inter-channel distance equal to 70/530. The shaded region represents the typical range of permeability of carbon felt electrodes in RFBs.[70–72]

To rationalize these tradeoffs, we predicted the spatial variation of superficial velocity $\vec{u}_s$ by solving Darcy's law subject to volume conservation: $\vec{u}_s = -(k_h/\mu)\nabla p$ and $\nabla \cdot \vec{u}_s = 0$, where $p$ and $\mu$ are pressure and dynamic viscosity, respectively. Here, hydraulic permeability $k_h$ depends on the phase associated with a given position: (1) micro-porous intercalating regions use the hydraulic permeability of unpatterned porous electrode material $k_h^{pe}$, whereas (2) macro-porous eµ-IDFF regions use a hydraulic permeability $k_h^{ch}$ that reproduces the axial-pressure/mean-velocity relationship resultant from Poiseuille flow through a given channel's cross-section. Our approach is an extension of Darcy-Darcy formulations that are routinely used to model flow through fractured, micro-porous rock.[73]

Channel width $w$, inter-channel distance $s$, and the gap between each channel's tip and the electrode's edge $g$ were varied to study their effects on the uniformity of the streamlines produced by a given eµ-IDFF design. Assuming $k_h^{pe} \approx 0.3$ µm², streamlines approach a perpendicular orientation to channel surfaces (i.e., transverse flow) for microchannels wider than 50 µm. Further, the model predicts more uniform flow through the porous electrode region between two adjacent microchannels with increasing $w$ and decreasing $s$ (Fig. 2e). Also, $g$ that is too large results in dead zones near channel tips, whereas $g$ that is too small suppresses transverse flow through the porous electrode (see SI and Figs. S3, S4). To understand the impact of the experimental variations of $k_h^{pe}$ (see Table 1) with various $k_h^{pe}$, we simulated the streamlines of the eµ-IDFF design sought later experimentally ($w$ = 70 µm, $s$ = 530 µm, and $g$ = 750 µm) that limits active material loss to approximately 11%. While low enough $k_h^{pe}$ values produce transverse flow, flow parallel to channels results from increasing $k_h^{pe}$ to values that are too large. The resulting apparent permeability values predicted among the different electrode samples agree well with experimental values as shown in Fig. 2g.

A simplified quasi-1D analytical model (see SI) also reveals that the transitions observed when varying channel width and when varying $k_h^{pe}$ are captured by a critical value of a common dimensionless parameter $\Xi = k_h^{pe} L^2/(w^3 s)$ that represents the characteristic ratio of channel hydraulic resistance to porous-electrode hydraulic resistance. Designs with $\Xi < 1$ assure that streamlines are routed between channels within porous electrode regions, rather than parallel to them, because of the finite transverse velocity produced across a given channel's entire length (Fig. S6a). This criterion explains why microchannel eµ-IDFFs produce effective flow-through intercalation electrodes, though microchannel eµ-IDFFs produce parallel flow for the high-permeability electrodes used commonly in RFBs. To produce the same $\Xi$ value, an RFB electrode with 10 µm² permeability must have three-fold larger channel width than an FDI electrode with 0.3 µm² permeability (assuming a common electrode length $L$ and inter-channel distance $s$), since $w_{FDI}/w_{RFB} = \left(k_{h,FDI}^{pe}/k_{h,RFB}^{pe}\right)^{1/3}$ is satisfied for FDI and RFB designs subject to such conditions. Thus, the use of microchannel eµ-IDFFs is not a simple extension of the past use of IDFFs in RFBs, and their ability to facilitate transverse flow is enabled by the patently low permeability of the porous material in which they are embedded. We also used the quasi-1D model to find a closed-form expression for eμ-IDFF apparent permeability (see SI), which is in good agreement with our numerical results using a channel width of 70 μm (Fig. 2g). By using the quasi-1D model, Fig. 2g also shows that to produce a normalized apparent permeability exceeding unity microchannels must exceed a certain critical width $w_{cr}$ that decreases with decreasing porous electrode permeability (Fig. S6c).

Understanding of laser/electrode interactions is critical to fabricate eμ-IDFFs that approach felicitous designs and that maintain the integrity of the underlying electrode material. We characterized the microstructure and composition of electrode material and the morphology of microchannels by engraving separate microchannels at various laser powers. At high laser power density we observed heat-affected zones (HAZs) that extend away from channel edges. Within such regions electrode material shows a permanent change in color from dark green to black (Fig. 3b) that is attributed to the structural decomposition of NiHCF, as evidenced by the disappearance of the C≡N peak at ~2160 cm$^{-1}$ in local Raman spectra at channel walls (Fig. 3d). Such changes suggest that the temperature at the laser-irradiated surface exceeds the onset of the NiHCF decomposition event at 395°C, as observed from thermogravimetric analysis (Fig. 3e). Further, the coarsening of electrode material in HAZs is likely due to the melting of PVDF at approximately 160°C.[74]

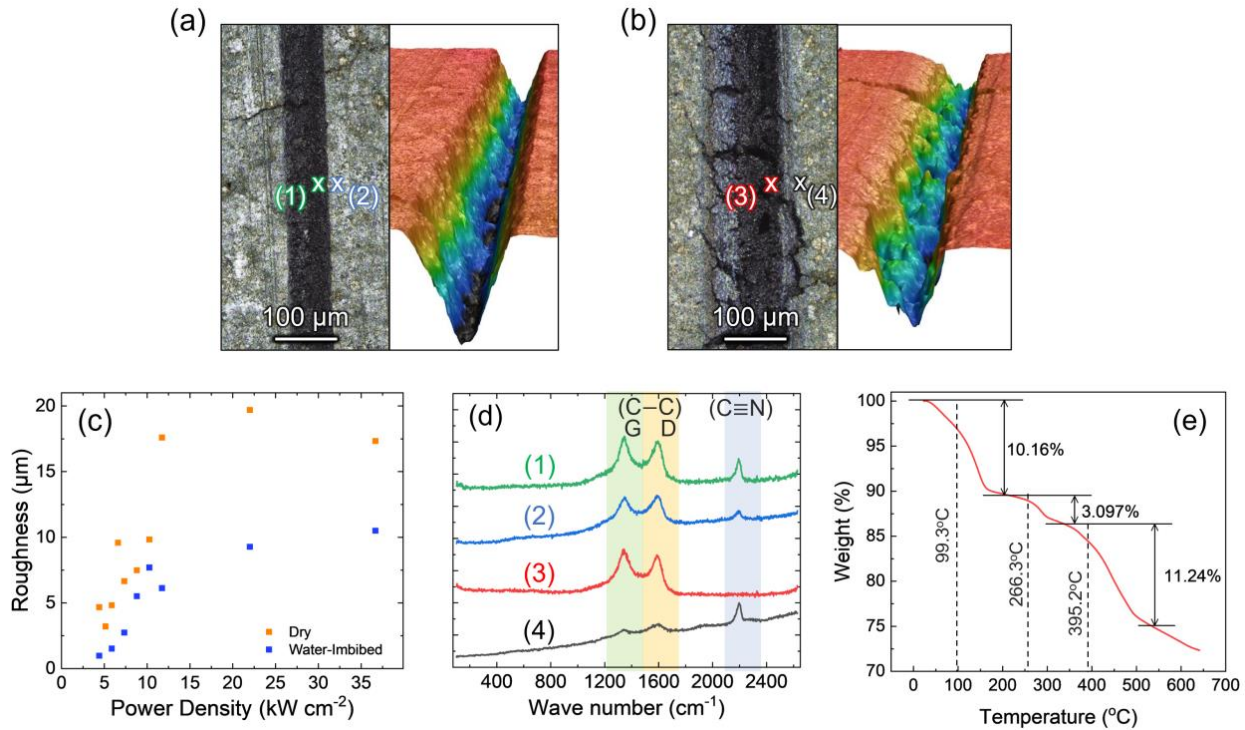

**Figure 3**. Optical and 3D profilometry images of microchannels engraved using (a) a water-imbibed electrode and (b) a dry electrode. (c) The root-mean-square (RMS) surface roughness of the channels made on dry and water-imbibed electrodes at different power densities. (d) Raman spectra measured inside and outside of channels made by water-imbibed engraving [(1) and (2) in (a)] and dry engraving [(3) and (4) in (b)]. Peaks at 1350 cm$^{-1}$, 1580 cm$^{-1}$, and 2300 cm$^{-1}$ are signatures of the D- and G-bands of C-C vibration and of C≡N bonds. (e) Thermogravimetric analysis of a calendered electrode showing water evaporation below 300°C and the decomposition of NiHCF at 395°C.

To mitigate these effects, we also performed laser micromachining on electrodes imbibed with water, inspired by water-assisted laser micromachining of non-porous materials.[75,76] Not only did water-imbibed engraving minimize HAZs (Fig. 3a), it also created smoother channels that are evidenced by smaller root-mean-square surface roughness within channels (Figs. 3c). These effects result from the ability of water to absorb some of the heat and mitigate the temperature rise in the electrode material surrounding the irradiated region and to prevent PVDF coarsening at 160°C and NiHCF decomposition at 395°C. The smaller thermal diffusivity[77] $\alpha$ of liquid water (0.15 mm$^2$/s) relative to air (20 mm$^2$/s) suggests ten-fold smaller penetration depth $\delta_P$ into HAZs when using water-imbibed engraving ($\delta_P \sim \sqrt{\alpha \tau}$ over time-scale $\tau$), in addition to the potential for thermal protection of HAZs due to the enthalpy of water vaporization.[†] Using these results, we determined specific laser settings to achieve the desired eµ-IDFF dimensions while minimizing HAZs (see SI). While the targeted eµ-IDFF design had microchannels with a 70 µm wide by 200 µm deep rectangular cross-section (Fig. 2d), their fabrication proved challenging due to the laser beam's tendency to engrave side walls that deviate from vertical orientation (Figs. 3a,b), likely as a result of its Gaussian intensity distribution. Therefore, we engraved microchannels with similar cross-sectional area to the felicitous design (14,000 µm$^2$) to produce similar hydraulic resistance, the dimensional characterization of which is included in the SI.

### *Desalination Experiments using eµ-IDFFs*

We performed desalination experiments using electrodes fabricated with eµ-IDFFs in a symmetric FDI flow cell (Fig. 4a and Fig. S13). During each desalination half-cycle one electrode captures cations to produce desalinated water while the other releases intercalated cations to produce brine. At the end of each half-cycle the former electrode approaches 100% state-of-charge (SOC) while the latter approaches 0% SOC. The applied current is then switched off and the FDI cell enters an open-circuit (OC) mode at $t_1$. To enable continuous desalination via the switching of electrical current we used a tubing system with four fluidic valves to switch inlet and outlet streams,[26] the effect of which is illustrated in Fig. 4b. A pause period ($\Delta t_{pause} = t_3 - t_2$) is then used between the time instants at which inlet ($t_2$) and outlet ($t_3$) switching events occur to allow diluate and brine within the FDI cell to evacuate into their respective reservoirs. This pause period minimizes effluent intermixing but cannot eliminate it completely. Typical real-time variations of cell potential, current, and diluate concentration during a desalination experiment are shown in Fig. 4c. During a given half-cycle, cell potential increases to 0.4 V or decreases to -0.4 V depending on the direction of the applied current.

---

[†]At the ~100 ns time-scale for laser irradiation, water evaporation occurs at higher temperatures and pressures than at equilibrium,[78] thus requiring further investigation to quantify the relative impacts of water's sensible and latent heating on HAZs during water-imbibed engraving.

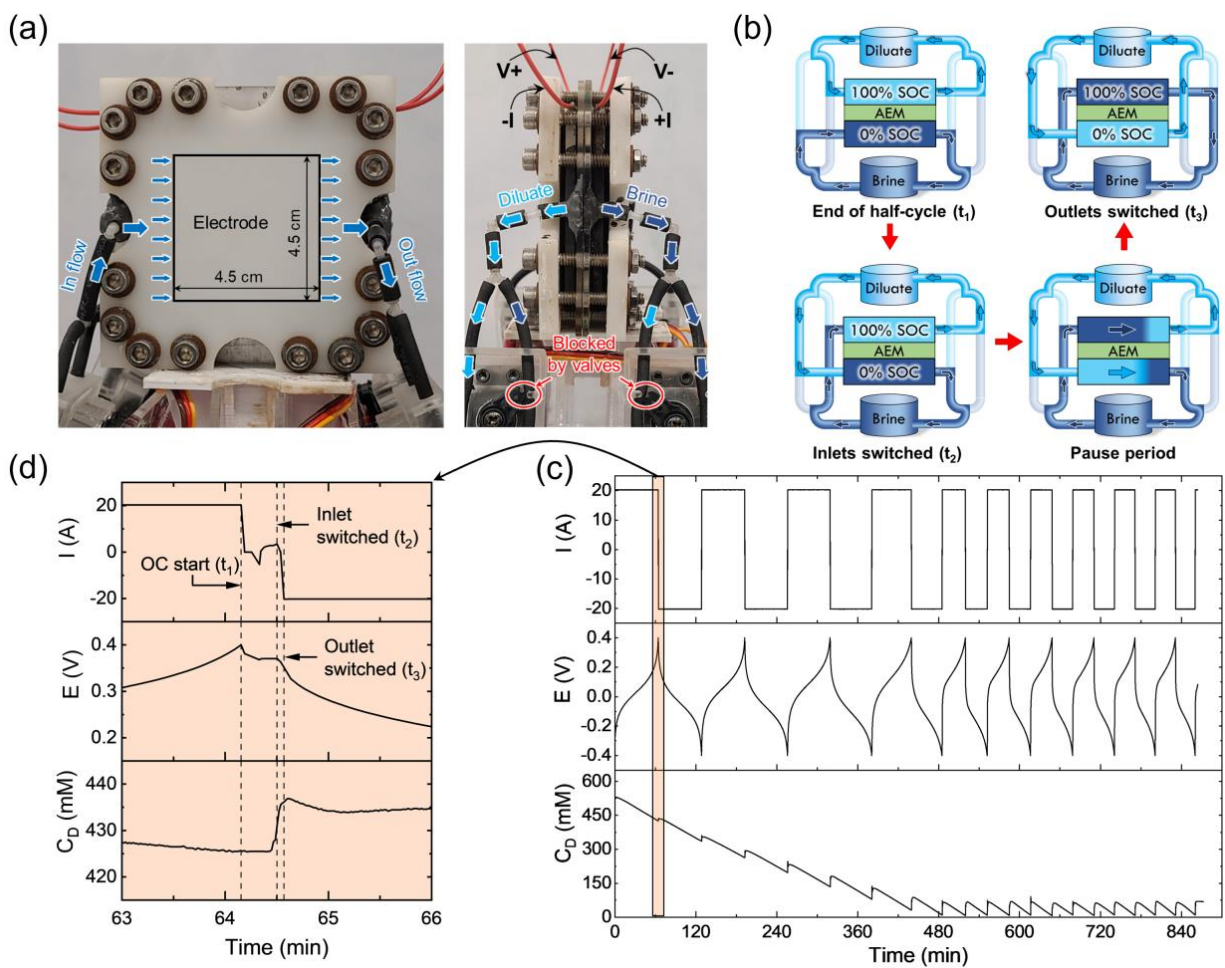

**Figure 4**. (a) Photos of the FDI flow cell with flow directions and applied current/voltage annotated, (b) schematic of the flow cell with fluid recirculation system during valve switching, with electrodes having disparate SOC. A half-cycle ends at $t_1$ and the applied current is terminated to produce open-circuit conditions, during which potential-controlled electrochemical impedance spectroscopy is performed. At $t_2$ the two inlet valves are switched. A pausing time $\Delta t_{pause}$ is used to let diluate and brine within the FDI cell evacuate into the correct reservoir. After $\Delta t_{pause}$ the outlet valves are switched at $t_3$. A similar sequence is used to switch the system back to its original state (i.e., the state at $t_1$) after cycling with oppositely signed current. (c) Time variation of FDI cell potential, applied current, and diluate-reservoir concentration from a desalination experiment using H-pair electrodes (21 mg cm$^{-2}$) with 1 mA cm$^{-2}$ applied current density. The shaded region is zoomed in and shown in (d).

The salt concentration in the diluate reservoir decreases continuously during each half-cycle (Fig. 4c) and rapidly increases for a short period of time after valve switching at the end of each half-cycle (Fig. 4d) due to effluent intermixing.[26] As a result such jumps in diluate-reservoir concentration become more pronounced as the difference in salt concentration between reservoirs becomes larger. The salt-concentration increase that results from such mixing eventually becomes large enough to match the salt removal obtained due to the applied current, thus limiting overall salt removal of the multi-cycle desalination process.

For feed water having salt concentration $C_F$ near seawater salinity, desalination experiments were conducted at 5 mL min$^{-1}$ flow rate with 1 mA cm$^{-2}$ current density. As shown in Fig. 5a, salt removal increases with increasing NiHCF loading from nearly 90% for L-pair electrodes (15 mg cm$^{-2}$) to 95% for M-pair electrodes (19 mg m$^{-2}$) to 96% for H-pair electrodes (21 mg cm$^{-2}$). Diluate feed concentration also decreased from 496 mM NaCl respectively to 102 mM, 23.7 mM, and 19.1 mM as a result of desalination using these electrode pairs. These results confirm our hypothesis that scaled-up FDI cells with high-capacity electrodes increase salt removal: increasing NiHCF loading by 40% from 15 mg cm$^{-2}$ to 21 mg cm$^{-2}$ reduced effluent concentration from brackish water salinity (102 mM) to near-potable salinity (19.1 mM), which is also the lowest salinity produced from such high-salinity feeds using CDI or FDI to our knowledge.

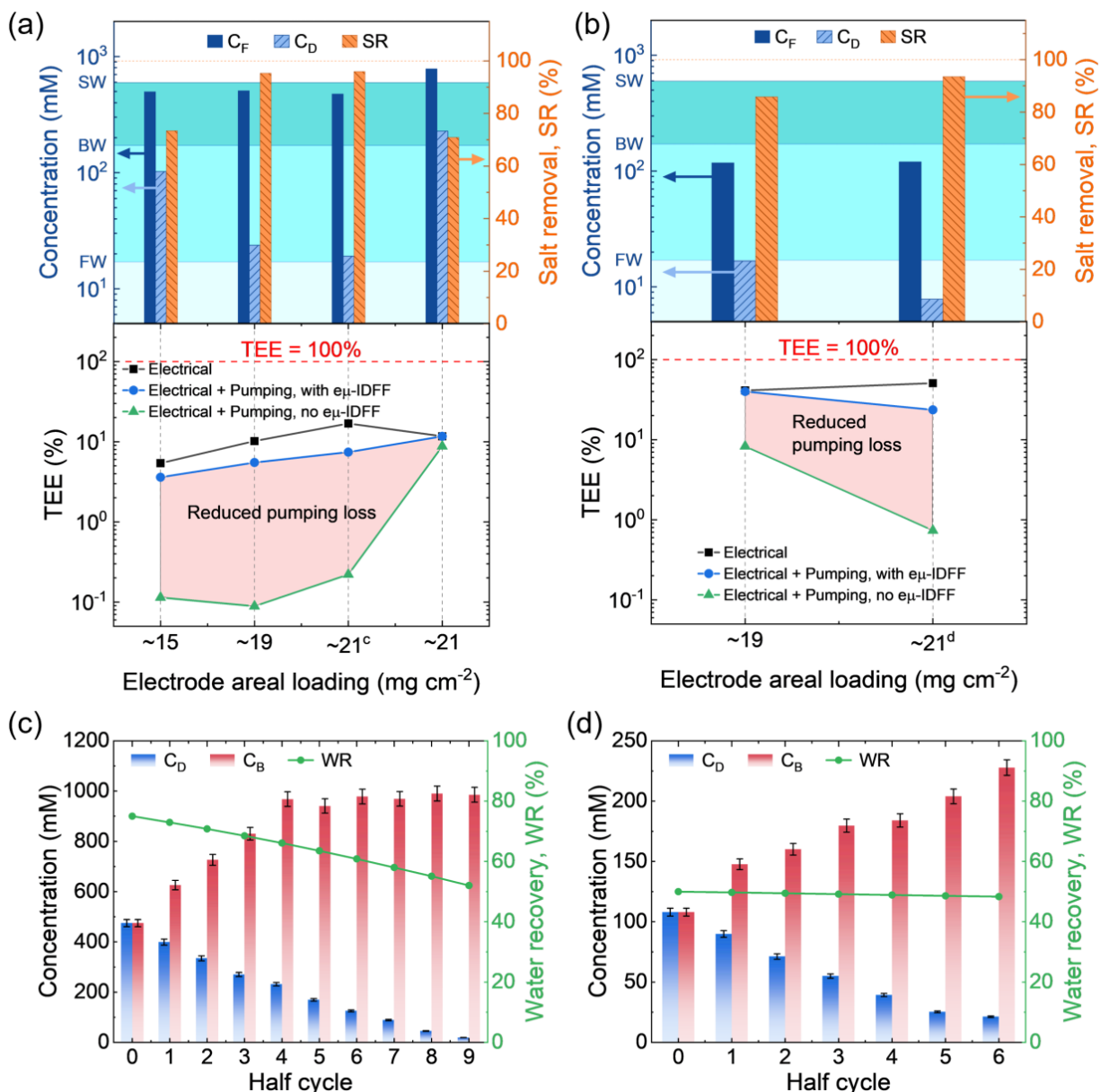

**Figure 5**. (a) Desalination of feed water with near seawater level salinity (SW, ~35 g L$^{-1}$ or 599 mM NaCl) down to brackish water level salinity (BW, <10 g L$^{-1}$ or 171 mM NaCl) and freshwater level salinity (FW, ~1 g L$^{-1}$ or 17 mM NaCl) and their corresponding thermodynamic energy efficiencies (TEEs). (b) Desalination of brackish water down to freshwater level salinity using different electrode loading levels and the resulting TEEs. The cycle dependence of diluate and brine reservoir concentrations for brackish and seawater experiments are shown in (c) and (d), respectively. The following symbols are used throughout: $C_F$ for feed concentration, $C_D$ for diluate concentration, $C_B$ for brine concentration, SR for salt removal, and WR for water recovery. Cyclic-specific data for all six experiments is provided in the SI (Fig. S14, S15).

Figures 5a,b show the corresponding thermodynamic energy efficiency (TEE) of these experiments considering (i) measured electrical input alone, (ii) measured electrical input and pumping energy calculated using the measured apparent permeability (with eµ-IDFF), and (iii) electrical input and pumping energy calculated using porous-electrode permeability (no eµ-IDFF). Using such dense and highly loaded electrodes without eµ-IDFFs is accompanied by infeasibly large pumping energy that decreases overall TEE. When considering only electrical energy input, the respective TEEs for L-, M-, and H-pair electrodes were 5.4%, 10.2%, and 17.0%. However, these respective TEEs diminished dramatically to 0.11%, 0.09%, and 0.22% when taking into account pumping energy without eµ-IDFFs. By engraving eµ-IDFFs into these electrodes, TEE was improved to 3.6%, 5.5%, and 7.4% due to 30- to 120-fold increases in their apparent permeabilities. These TEEs are at parity with past FDI experiments using intercalation materials, despite feed water salinity in such work being limited to 25 mM or less.[19] The pumping-energy saved by using eµ-IDFFs (shaded regions in Fig. 5a) indicates the essential role that eµ-IDFFs provide in enhancing FDI TEE using flow-through electrodes.

We also carried out desalination experiments on hypersaline brine using 21 mg cm$^{-2}$ H-pair electrodes. These experiments were able to use decreased flow rate (1 mL min$^{-1}$) and increased current density (8.5 mA cm$^{-2}$), owing to the decreased SOC gradients and ohmic resistance expected for such brines. We observed a diluate concentration decrease of 71% from 781 mM initially to 227 mM as a result (Fig. 5a), demonstrating for the first time the potential of FDI using cation intercalation electrodes to treat concentrated brines. Furthermore, this level of salt removal was achieved with 11.7% TEE irrespective of whether pumping energy was included or not. The insignificance of pumping energy in this context can be understood in the following manner. $E_{pump}$ is the product of hydraulic resistance $R_h$, the square of volumetric flowrate $Q$, and desalination batch time $\Delta t$: $E_{pump} = R_h Q^2 \Delta t$, where $R_h$ is proportional to overall electrode length $L$ and is inversely proportional to the apparent hydraulic permeability $k_h^{app}$ according to Darcy's law ($R_h = \mu L/(k_h^{app} A)$, where $A$ is the area of the electrode's inlet plane). Therefore, pumping energy at higher current density decreases as desalination batch time shortens, and consequently its contribution to TEE losses becomes less significant in that limit. This finding suggests that using electrodes tailored to operate at high current density and low flow rate could further decrease pumping energy and improve overall TEE.

For brackish water desalination, the use of eµ-IDFFs results in exceptional TEE. Two experiments were conducted in which M-pair (19 mg cm$^{-2}$) and H-pair (21 mg cm$^{-2}$) electrodes were used to desalinate feed water with ~100 mM NaCl at 1 mA cm$^{-2}$. As shown in Fig. 5b, salt concentration decreased from 118mM to 16.7 mM for M-pair and to 7.8 mM for H-pair, both of which are below the salinity of fresh water (17 mM). With eµ-IDFFs the TEE for M-pair electrodes increased from 8.3% to 40.0% and that of H-pair electrodes increased from 0.7% to 23.6%. The higher TEE of the former case is due to the use of a smaller flowrate (1 mL min$^{-1}$ compared to 5 mL min$^{-1}$ in the latter case). These TEEs approach that of a single stage RO system without energy recovery, according to recent analysis.[79]

We also find that eµ-IDFFs increase TEE not only by improving electrode hydraulic permeability but also by maximizing active-material capacity utilization. Figure

S16 shows the specific capacity versus cycle number obtained from desalination experiments using the L-, M-, and H-pair electrodes with eµ-IDFFs, along with that of an unpatterned electrode pair. In all cases a seawater-salinity feed was used. The electrodes with eµ-IDFFs consistently show specific capacities greater than 55 mAh g$^{-1}$, whereas the unpatterned electrodes show a maximum specific capacity of 47 mAh g$^{-1}$ only for the first half-cycle and subsequently show steadily decreasing capacity. This effect may be a result of dead volume within porous electrodes that is formed during calendering, where an initially interconnected porous network (~60% porosity) is compressed into a dense microstructure having approximately 40% porosity. As the porous network collapses, the solid constituents fill micropores to form isolated pockets to which fluid delivery is impeded during desalination experiments, potentially causing insufficient salt delivery to intercalation material and thus reducing capacity. Engraving electrodes with eµ-IDFFs likely opens such isolated pores to enable intercalation material to access fresh salt solution.

Water recovery (WR), defined as desalinated water volume divided by total feed-water volume, was determined at the end of each desalination half-cycle. While terminal WR exceeded 50% for all experiments (Figs. 5c,d), seawater and hypersaline brine feeds showed gradually decreasing WR with increasing salt removal (Fig. 5c), thus requiring compensation by using increased initial diluate volume to achieve 50% terminal WR. From this data we determined water transport rates between diluate and brine as high as 0.16 L m$^{-2}$ h$^{-1}$ (Table S5). Following past theoretical analysis for osmosis/electroosmosis through ion-exchange membranes (IEMs),[80–82] we predict a maximum water transport rate through Neosepta AMX as 0.20 and 0.25 L m$^{-2}$ h$^{-1}$ when using respective salt concentrations of 496 mM and 781 mM and assuming 40 vol.% water therein.  These values are consistent with previous measurements on commercially available membranes,[83] ED stacks,[84] and RFBs[85] (0.24 – 2.46 L m$^{-2}$ h$^{-1}$), making water flux through IEMs a probable mechanism of WR loss among other factors.

Despite the promising TEEs achieved in comparison with prior CDI and FDI literature, our results indicate significant room for improvement relative to the thermodynamic limit of energy consumption, especially for high-salinity feeds. For this reason we analyzed process charge efficiency[86] (CE), defined as the ratio of cationic charge desalinated to electrical charge transferred.  In general, CE was shown to decrease with increasing feedwater salinity.  Despite brackish feeds with less than 100 mM NaCl exhibiting nearly ideal CE (Fig. S15), we observed a decrease of CE to ~50% when using seawater salinity feeds (Fig. S14).  To isolate CE losses due to effluent intermixing (a type of 'flow efficiency' loss mechanism[87,88]) we performed separate desalination experiments for a single half-cycle (i.e., experiments without valve or current switching) that used H-pair electrodes (21 mg cm$^{-2}$) with various feed concentrations.  These experiments showed CE dropping linearly from 91.7% to around 50% as feed concentration increased from 38 mM to more than 500 mM (Fig. S17a). Since our electrodes exhibited excellent cyclability with no decrease in specific capacity after 90 cycles (Fig. S17b), this concentration-dependent CE likely results from mechanisms other than electrode degradation, such as gas-evolution/-consumption reactions or the loss of membrane permselectivity.  Further, Fig. 2f shows that, even though the chosen eμ-IDFF generates streamlines that are perpendicular to microchannels, streamlines are concentrated near electrodes edges, exhibiting a degree of non-uniformity in fluid distribution. In addition, dead zones could exist within the electrode microstructure as a result of its heterogeneous nature, and side reactions occurring because of salt depletion within dead zones would lead to lower CE. On the other hand, selectivity losses are ubiquitous in membrane processes such as electrodialysis and ion concentration polarization.[89] When the counter-ion concentration around an IEM exceeds the concentration of the fixed-charge groups within the IEM, co-ions can penetrate into the IEM to cause co-ion leakage that acts against deionization.[89,90] Hence, IEM permselectivity likely contributes to the CE losses observed for high-salinity feeds.

**Conclusions**

In this work we fabricated three pairs of high areal-loading electrodes (L: 15 mg cm$^2$, M: 19 mg cm$^2$, and H: 21 mg cm$^2$) that were embedded with interdigitated microchannels, and we desalinated water using them in a symmetric FDI flow cell with feed water recirculation. These electrodes are shown to reduce salt concentration from seawater salinity (496 mM NaCl) down to brackish water salinity (102 mM for L-pair), and to near freshwater salinity (23.7mM for M-pair and 19.1 mM for H-pair). The overall TEEs of these three experiments were 3.6%, 5.5%, and 7.4%, respectively, which are similar to past FDI studies that used brackish water feeds. H-pair electrodes were also used to treat hypersaline influent with 781 mM NaCl and were able to produce effluent of 227 mM at 11.7% TEE. This demonstrates for the first time the potential of symmetric FDI to treat concentrated brines. For brackish water influent of 118 mM NaCl, M- and H-pair electrodes comfortably produced effluents with freshwater salinities (16.7 mM and 7.8 mM) at TEEs of 40.0% and 23.6%, respectively.  Moreover, these electrodes exhibited outstanding cyclability with unnoticeable drops in specific capacity after 90 cycles.  Despite this, mild yellowing of the AEM was observed after such experiments, the effect of which we postulate as mild dehydrochlorination caused by weak Faradaic side reactions at the electrode/AEM interface (see SI, Fig. S18).  This study thus demonstrates experimentally for the first time that FDI using scaled-up, high-capacity electrodes can desalinate seawater-level salinity to near freshwater salinity, and it can substantially decrease the salinity of hypersaline brines by resolving the issue of pumping pressure/energy that is inherent to flow-through electrodes.

Factors contributing to energy efficiency losses are also identified. First, water transport between diluate and brine was observed as high as 0.15 L m$^{-2}$ h$^{-1}$, but water recovery exceeding 50% was still obtained by using excess water initially in the diluate reservoir.  Second, the charge efficiency in the first half-cycle of desalination experiments decreased from 91.7% to around 50% when feed concentration was varied from 38 mM to greater than 500 mM.  Chemical, physical, and electrochemical interactions between porous electrode material, membrane material, and feed/effluent solution therefore require further investigation to increase process efficiency, water recovery, and water productivity.

We have demonstrated that minimizing the longitudinal hydraulic resistance within microchannels relative to the transverse hydraulic resistance through intervening porous-electrode material is critical to the effective use of eμ-IDFFs, thus demonstrating the importance of flow-field co-design with the porous electrodes in which they are embedded.  Because electrode permeability scales with the square of particle size, eμ-IDFFs are therefore likely to have broad impact on flow-through electrodes that incorporate fine-scale particles. Therefore, eµ-IDFFs could also find use in flow batteries, fuel cells, and $CO_2$ conversion electrodes that have employed nanomaterial catalysts[91–94] or conductive additives,[92] in addition to selective electrochemical removal and recovery processes. Water imbibition can also be used to protect other porous electrode materials during laser micromachining for Li-ion batteries[95] and micro-supercapacitors.[96]

**Experimental Methods**

*Synthesis of NiHCF Nanoparticles*

Nickel hexacyanoferrate (NiHCF) nanoparticles – a type of Prussian Blue analogue intercalation material – were prepared as in our previous work.[26] Two solutions of 0.1 M $K_3Fe(CN)_6$ and 0.2 M $NiCl_2$ with 1:1 volume ratio were added dropwise into a beaker containing deionized (DI) water and stirred vigorously at room temperature to obtain a suspension of NiHCF nanoparticles, which was then sonicated for 30 mins and aged overnight. The nanoparticles were subsequently collected by centrifugation and were rinsed with DI water and ethanol, followed by drying under vacuum at 80°C.

*Electrode Fabrication*

Porous electrodes were made from 80:5:15 wt.% of NiHCF, Ketjen black (KB) conductive additive (EC-600JB), and polyvinylidene fluoride (PVDF) binder (Solvay Solef 5130), respectively. The mixture of NiHCF particles with KB was dry ground in a vortex mill using 5 mm steel balls (Ultra Turra-X, IKA) at 6000 rpm for 30 minutes to obtain a fine, homogeneous powder. PVDF was dissolved in N-methyl-2-pyrrolidone (NMP, Sigma Aldrich) to obtain a viscous, transparent liquid. The NiHCF and KB powder mixture was added to this solution resulting in 3 mL of NMP per 1 g of solid material. These components were homogenized in a planetary mixer (Thinky, ARE-310) for 30 minutes. The resulting slurry was cast at thicknesses of 1.1 mm for L-pair electrodes, 1.3 mm for M-pair electrodes, and 1.4 mm for H-pair electrodes onto graphite-foil (Ceramaterials) current-collectors by using a doctor blade and film applicator (Elcometer 4340). We note here that we typically use 1 g of solid materials for one batch of slurry, and 1.4mm thick is the maximum thickness we could get to produce a nice 4.5×4.5cm electrode after removing the material near the edges, which is prone to crack during calendaring. Increasing the cast thickness further may cause the electrode area to be smaller than 4.5×4.5cm with the prepared slurry. While making more slurry by using more solid materials and NMP is feasible, we did not do that since optimizing electrode mass loading is not the main focus of this study. Immediately after casting, electrodes were immersed in an alkaline bath with pH=12 at 85°C for approximately 1 minute. These electrodes were then rinsed with DI water and subsequently dried in a fume hood to produce thickness of greater than 400 μm. They were then calendered down to ~200 μm using a roll press (MTI-XTL) and were subsequently engraved with an interdigitated pattern over a 45 × 45 mm area. This engraved area was then cut with scissors to obtain final electrodes.

*Porosity and Permeability Determination for Electrodes*

The porosity of each electrode $\varepsilon$ was calculated from the electrode's measured density $\rho_e$ and the mass-averaged density $\rho_c$ of its constituents (2.0 g cc$^{-1}$ for NiHCF,[33] 2.0 g cc$^{-1}$ for Ketjen Black,[97] and 1.75 g cc$^{-1}$ for PVDF binder[98]) as $\varepsilon = 1 - \rho_e/\rho_c$. The permeability of each electrode with thickness $t_e$ was measured using a gravity-driven apparatus[28] wherein a constant pressure head $\Delta p$ (in units of Pa) was applied via a column of water to the FDI cell containing the targeted electrode. The amount of water permeating through the electrode over a certain period of time was used to calculate the average volumetric flow rate $Q$ in units of m$^3$ s$^{-1}$. The permeability $k_h$ in units of m$^2$ was then calculated from Darcy's law: $Q = \Delta p k_h W t_e / L \mu$, where $\mu$ is water's dynamic viscosity in Pa-s, and $W$ and $L$ are sample width and length, respectively.

*Electrode Engraving by Laser Micromachining and Channel Characterization*

eμ-IDFFs were engraved using a Trotec Speedy Flexx 400 laser. Immediately prior to water-imbibed engraving, dry electrodes were first soaked in DI water and subsequently wiped a Kimwipe to remove excess water. In all instances electrode engraving was conducted using ventilation to avoid human exposure to laser ablation products (e.g., HCN and HF). Optical profilometry (Keyence VK-X1000), Raman spectroscopy (Nanophoton Raman 11), and thermogravimetric analysis were used to characterize the composition and thermal stability of electrode materials in heat affected zones.

*Symmetric FDI Cell Setup and Fluid Recirculation System*

After laser engraving, electrodes with 45 × 45 mm size were cycled at a C-rate of C/10 in a three-electrode setup in 500 mM NaCl with an Ag/AgCl reference electrode (0.197 V vs. SHE) and a potentiostat (Biologic VMP-3). This step was used to remove Na$^+$ and K$^+$ present in the NiHCF crystal lattice during synthesis. Before being assembled in the FDI cell, one electrode was reduced to ~0% state-of-charge (SOC) (0.1 V vs. Ag/AgCl) and the other one was oxidized to ~100% SOC (0.6 V vs. Ag/AgCl) by chronopotentiometry in a three-electrodes set up.

The FDI cell used in this study (Fig. 4a) is a scaled-up version of the cell used in our previous work[26] that contains two 45 × 45 mm electrodes separated by an anion-exchange membrane (Neosepta AMX). In all experiments a potentiostat (Biologic VMP-3) was used to control electrochemical cycling. A four-probe connection scheme (i.e., a Kelvin sensing measurement) was used to mitigate contact resistance between the potentiostat and the flow cell, as illustrated by distinct leads for sensing voltage and driving current in Fig. 4a. Salt concentrations used to calculate salt removal were determined by ion chromatography performed using a 930 Compact IC system from Metrohm with a Metrosep C4 – 150/4.0 cation column. Real-time conductivity of the reservoirs was collected using the Conduino system.[99]

*Performance metrics*

*Thermodynamic energy efficiency:*

TEE for a desalination experiment was calculated as:

$$TEE = \frac{SEC_{min}}{SEC} \times 100$$

The specific energy consumption $SEC$ (kJ mol$^{-1}$) was calculated by dividing the total electrical energy input to the moles of salt removed from diluate stream into brine stream $n_{salt}$, which was measured at the end of the experiment:

$$SEC = \frac{\int_{t_0}^{t_{end}} I(t)V(t)dt}{n_{salt}}$$

with $I(t)$ and $V(t)$ respectively being the applied current and full-cell potential at time $t$. The minimum specific energy consumption was calculated per mole of salt removed as:

$$SEC_{min} = \frac{W_{rev}}{n_{salt}}$$

Here, the reversible work of separation $W_{rev}$ was calculated using non-ideal activity coefficients due to the high-salinity feeds used in our experiments and the hypersaline brines produced from them:

$$W_{rev} = 2RT\left[V_D C_D \ln\left(f_\pm^{C_D} C_D\right) + V_B C_B \ln\left(f_\pm^{C_B} C_B\right) - (V_D + V_B)C_F \ln\left(f_\pm^{C_F} C_F\right)\right]$$

Here, $V_i$ and $C_i$ are respectively the volume of water and salt concentration of solution $i$, where $i = B, D$, and $F$ respectively refer to brine, diluate, and feed. The mean activity coefficient $f_\pm^C$ was estimated for a given salt concentration $C$ using literature data.[100]

*Charge efficiency*

Charge efficiency was calculated from the ratio of the moles of electrons transferred throughout an experiment $n_{charge}$ to the moles of salt removed $n_{salt}$:

$$CE = \frac{n_{salt}}{n_{charge}} \times 100\%$$

*Water recovery*

Water recovery was calculated at the end of each desalination half-cycle by using the volumes of the diluate $V_D$ and brine $V_B$ reservoirs, assuming that salt and liquid volume were conserved:

$$WR = \frac{V_D}{V_D + V_B} \times 100\%$$

*Water transport*

The water transport in each experiment was calculated as the difference between the volume of the diluate reservoir initially ($V_D^0$) and after the experiment finished ($V_D^{end}$), divided by the time duration of the experiment $\Delta t$ and the area of the anion exchange membrane $A_{AEM}$:

$$J_w = \frac{V_D^0 - V_D^{end}}{\Delta t A_{AEM}}$$

*Simulation of Flow through Electrodes with eμ-IDFFs*

Flow-through electrodes embedded with eμ-IDFFs were modeled using conservation of fluid volume with a Darcy-type rate law for superficial velocity $\vec{u}_s$, as described already ($\vec{u}_s = -(k_h/\mu)\nabla p$). For each $(x,y)$ location inside of a microchannel region we used $k_h(x,y) = k_h^{ch}$, where $k_h^{ch}$ was calculated based on the Boussinesq solution for Poiseuille flow through a channel of rectangular cross-section with $h$-by-$l$ size (see SI):[101]

$$k_h^{ch} = \frac{h^2}{12} - \frac{16h^3}{l\pi^5} \sum_{n=1}^{\infty} \frac{1}{(2n-1)^5} \frac{\cosh(\beta_n l) - 1}{\sinh(\beta_n l)} \qquad (3)$$

Otherwise $k_h(x,y) = k_h^{pe}$ was used. The numerical solution for $\vec{u}_s$, obtained via the finite volume method implemented in MATLAB and solved using an aggregation-based algebraic multigrid method,[102–104] was then used to calculate the stream function[105] $\psi$ via numerical integration:

$$\psi(x,y) = \psi_0 + \int_{(x_0,y_0)}^{(x,y)} u_{s,x} dy - u_{s,y} dx \qquad (4)$$

Equally spaced contours of $\psi$ were then used to determine streamlines.

**Acknowledgements**

The Expeditionary Energy Program of the US Office of Naval Research (Award no. N00014-22-1-2577), the Chemical, Bioengineering, Environmental and Transport Systems Division of the US National Science Foundation (Award no. 1931659), and the Department of Mechanical Science and Engineering at the University of Illinois at Urbana-Champaign (UIUC) supported this research. Thermogravimetric analysis, optical profilometry, and Raman spectroscopy were performed at the Materials Research Laboratory at UIUC. We thank Neil Pearse and Michael Dalton for the use of the Trotec Speedy Flexx 400 Laser in the Siebel Center for Design at UIUC. We thank Paolo Luzzatto-Fegiz for sharing the Conduino system used here. We thank Rob Roberts, Patrick Fahey, and Yarik Syzdek of Biologic for helpful conversations about Kelvin sensing and conductivity measurement by impedance. We thank Nouryon Chemicals for providing Ketjen black material, and we thank All Foils for providing thin Al foil for laser testing.

**Conflicts of Interest**

KCS, VQD, ERR, and ICL declare their filing of a related patent application.

2000).

# Supporting Information for

# "Embedded, micro-interdigitated flow fields in high areal-loading intercalation electrodes towards seawater desalination and beyond"

Vu Q. Do,[a] Erik R. Reale,[a] Irwin Loud IV,[a] Paul G. Rozzi,[a] Haosen Tan,[b] David A. Willis,[b] and Kyle C. Smith[a,c,d,e,*]

a. Department of Mechanical Science and Engineering,
University of Illinois at Urbana-Champaign, Urbana, IL, USA

b. Department of Mechanical Engineering,
Southern Methodist University, Dallas, TX, 75275

c. Department of Materials Science and Engineering,
University of Illinois at Urbana-Champaign, Urbana, IL, USA

d. Computational Science and Engineering Program,
University of Illinois at Urbana-Champaign, Urbana, IL, USA

e. Beckman Institute for Advanced Science and Technology,
University of Illinois at Urbana-Champaign, Urbana, IL, USA

*corresponding author: kcsmith@illinois.edu

**Table of Contents**

# Modeled Comparison of Desalination using Different Flow Fields

Using the flowing porous electrode theory implemented in Refs. 1,2 and using the same model parameters described therein, we conducted simulations to compare the salt removal dynamics of representative flow-by configurations (both flow behind and flow between) to a flow-through configuration as shown in Fig. S1. As we demonstrate using other modeling in the main text and later in the SI, appropriately designed interdigitated flow fields produce flow *through* electrodes, rather than flow *by* electrodes. In contrast, the serpentine and parallel flow-field designs that have been used in fuel cells and flow batteries previously do not primarily produce flow through porous electrodes. Hence, the purpose of the present porous electrode theory modeling is to illustrate the impact of flow-through versus flow-by electrodes on the electrochemical transport processes that occur when used in a symmetric FDI device for desalination.

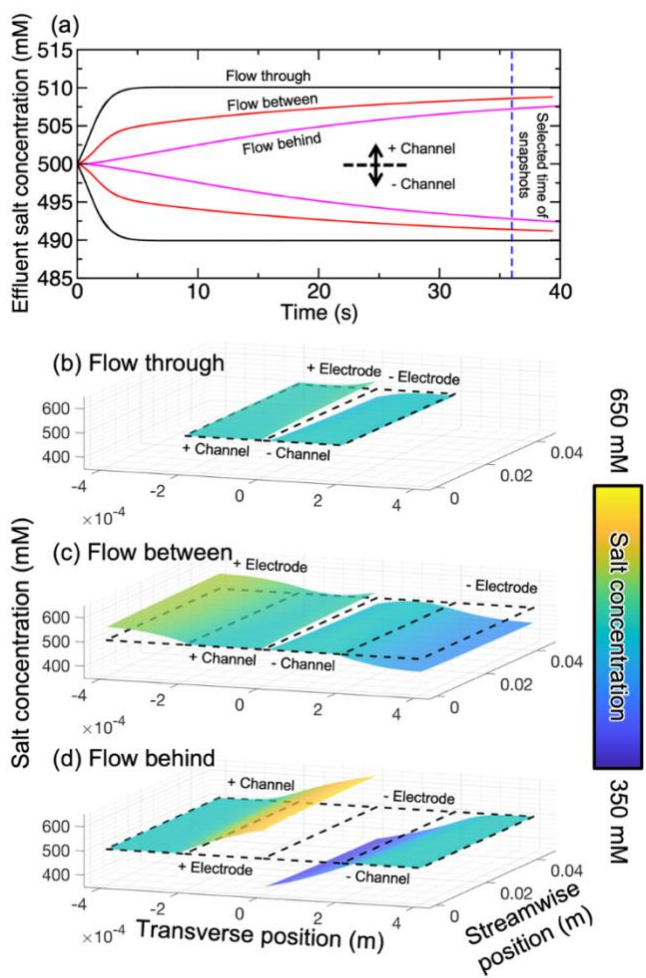

**Figure S1**: Simulated response at 4 mA cm$^{-2}$ applied current density for electrodes having 16 mg cm$^{-2}$ loading and 45 mm x 45 mm electrode area for flow-through and flow-by configurations. (a) Effluent salt concentration versus time. Spatial distribution of salt concentration for (b) flow-through, (c) flow-between, and (d) flow-behind configurations.

To this end Fig. S1b,c shows that in flow-by configurations salt is removed from solution that is stagnant, rather than from flowing solution. As a result, larger concentration differences persist in the transverse direction across flow-by electrodes (Figs. S1b,c) in comparison with flow-through electrodes (Fig. S1d). The impact of these effects is reduced desalination efficiency, because of the more lagged response of the effluent concentration produced by flow-by configurations relative to flow-through configurations (Fig. S1a). Hence, flow-by configurations, such as those engendered by serpentine and parallel flow fields, are likely to experience substantial losses to salt removal due to this effect. In contrast, the solution within electrodes is efficiently evacuated from the electrodes when a flow-through configuration is used (see Figure S1b).

## Asymptotic Scaling of Apparent Permeability

The flow rate through a channel due to a pressure difference between the inlet and outlet plane $\Delta p = p_{in} - p_{out}$ (Fig. S2a) can be expressed using Darcy's law if the hydraulic resistance within channels is neglected (in all other modeling that follows after this section – both numerical and analytical – we relax this simplifying assumption):

$$Q = Q_1 + Q_2 = 2Q_1 = 2\frac{k_h^{pe}}{\mu}\frac{\Delta p}{s} A_{in} = 2\frac{k_h^{pe}}{\mu}\frac{\Delta p}{s} hL_{ch} \qquad (*)$$

Considering the same section of the electrode, Darcy's law also relates the flow rate through this section with its apparent permeability $k_h^{app}$ and the pressure drop $\Delta p$, as illustrated in Fig. S2b:

$$Q = \frac{k_h^{app}}{\mu}\frac{\Delta p}{L_e} A_{in} = \frac{k_h^{app}}{\mu}\frac{\Delta p}{L_e}[2(s+w)h] \qquad (**)$$

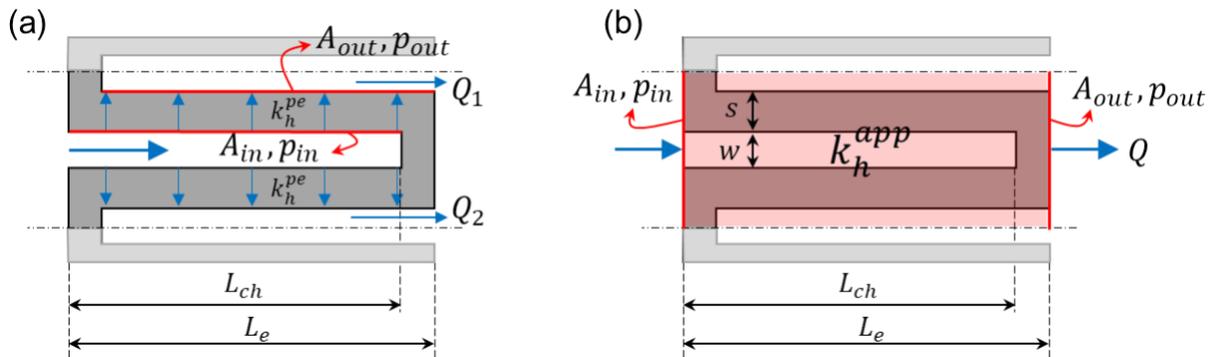

**Figure S2**: (a) Flow rate through a channel when considering the porous electrode hydraulic permeability $k_h^{pe}$. (b) Flow rate through a section of a patterned electrode when considering its apparent permeability $k_h^{app}$. Here, $w$ is channel width, $s$ is inter-channel spacing, $h$ is electrode thickness, $L_{ch}$ is channel length, and $L_e$ is electrode length. $A$ is area, $p$ is pressure, and $in$ and $out$ subscripts denote inlets and outlets.

Equating (1) and (2) we obtain the following scaling for the normalized apparent permeability in this limit of vanishing hydraulic resistance within channels:

$$k_{h,max}^{app,*} = \frac{k_{h,max}^{app}}{k_h^{pe}} = \frac{L_e L_{ch}}{s(s+w)}$$

Here, we denote this quantity as $k_{h,max}^{app,*}$ because an upper bound to apparent permeability is produced when the hydraulic resistance within channels is neglected. The above relation shows that if we keep constant electrode length $L_e$ and channel length $L_{ch}$, and if we also fix the coverage fraction of the electrode material for the IDFF (i.e, $s/w = $ constant), then the improvement in apparent hydraulic permeability due to IDFF embedding scales with the inverse square of channel width:

$$k_{h,max}^{app,*} = \frac{k_{h,max}^{app}}{k_h^{pe}} = \frac{L_e L_{ch}}{w^2 \alpha(\alpha+1)}$$

## Numerical Model for Flow-Through Electrodes Embedded with eμ-IDFFs

When designing an eμ-IDFF, there are three dimensions that we varied: channel width $w$, inter-channel distance $s$, and channel length $L_{ch}$. The last of these dimensions determines the gap $g$ between the end of the channel and the edge of the electrode. The objective of choosing the dimension of a given eμ-IDFF is to ensure even distribution of flow through the electrode microstructure while minimizing material loss. To model these effects, we implemented a finite-volume solver in MATLAB to simulate a Darcy-Darcy model for the 2D superficial velocity $\vec{u}_s$ inside porous electrodes that uses a permeabilty $k_h^{pe}$ within porous electrode regions and that uses a permeability $k_h^{ch}$ in channel regions. $k_h^{ch}$ was determined from the Boussinesq solution for Poiseuille flow in rectangular cross-section channel with $h$-by-$l$ size[4]:

$$Q = \frac{k_h^{ch} A}{\mu} \nabla p = \frac{hl\nabla p}{\mu}\left[\frac{h^2}{12} - \frac{16h^3}{l\pi^5}\sum_{n=1}^{\infty}\frac{1}{(2n-1)^5}\frac{\cosh(\beta_n l)-1}{\sinh(\beta_n l)}\right]$$

where $Q$ is flow rate, $p$ is pressure, and $A$ is area normal to flow in the channel. This results in the following expression for $k_h^{ch}$:

$$k_h^{ch} = \frac{h^2}{12} - \frac{16h^3}{l\pi^5}\sum_{n=1}^{\infty}\frac{1}{(2n-1)^5}\frac{\cosh(\beta_n l)-1}{\sinh(\beta_n l)} \qquad (S5)$$

We solve the discrete equations obtained using the Finite Volume Method (FVM) using an iterative linear solver based on the Aggregation Based Algebraic Multigrid (AGMG) method.[5–7]

When electrode permeability and porosity are kept constant, a large gap $g$ produces regions between channel tips and electrode edges where streamlines are absent, indicating that fluid does not flow through electrode material and instead bypasses those regions (Fig. S3). Hence, small $g$ is desirable to eliminate such dead zones. In addition, when $g$ is fixed at 0.75 mm, increasing microchannel width and decreasing channel spacing leads to more uniform flow distribution through the porous electrode regions (Fig. S4).

The differences between the predicted apparent permeability and the measured apparent permeability for the electrodes used in our experiments (Fig. 2g in the main text) arise due to a number of reasons. Firstly, we assume that the permeability within porous electrode regions is uniform, when in fact the fabricated electrodes possess a certain level of heterogeneity in their permeability. For example, we observed electrodes to possess initial thickness before calendaring that can vary as much as 50 μm between the thinnest and the thickest regions, that is converted to a relative error in mass loading of $\sigma_{rel}(\Gamma) = 13\%$. In addition, when calculating porosity, we assumed that graphite foil thickness is maintained at 100 μm after calendaring. If it were compressed by 50% during calendering, the relative error of measured electrode thickness is $\sigma_{rel}(w_e) = 28\%$. The porosity, $\varepsilon$, depends on the electrode's density, $\rho_e$. Hence, the relative error of these quantities is the same: $\sigma_{rel}(\varepsilon) = \sigma_{rel}(\rho_e) = \sqrt{\sigma_{rel}^2(\Gamma) + \sigma_{rel}^2(w_e)} \approx 31\%$. These estimates of error are larger than the deviation between our predictions and experimental measurements, explaining the origin of these discrepancies at least in part. In addition, the model treats eμ-IDFFs as an infinite periodic array of channels, such that finite size effects are neglected, despite their presence in experimental patterns.

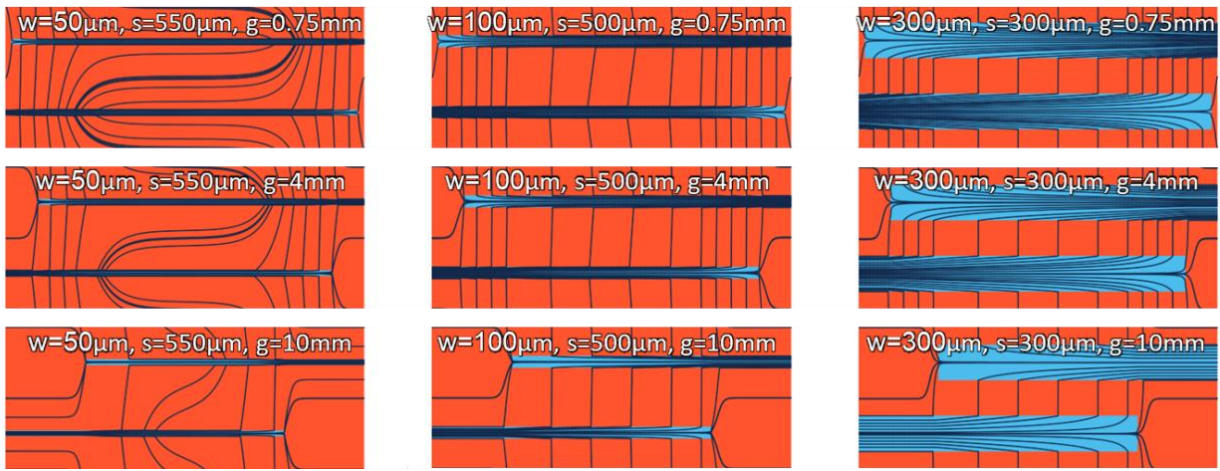

**Figure S3**. Streamlines of different interdigitated flow fields (eμ-IDFFs) with inter-channel distance $s$, channel width $w$, and gap between channel end and electrode edge $g$ as shown. The electrode permeability and porosity in all cases are 0.28 μm$^2$ and 45%. Vertical and horizontal scaling are set differently to ease streamline visualization.

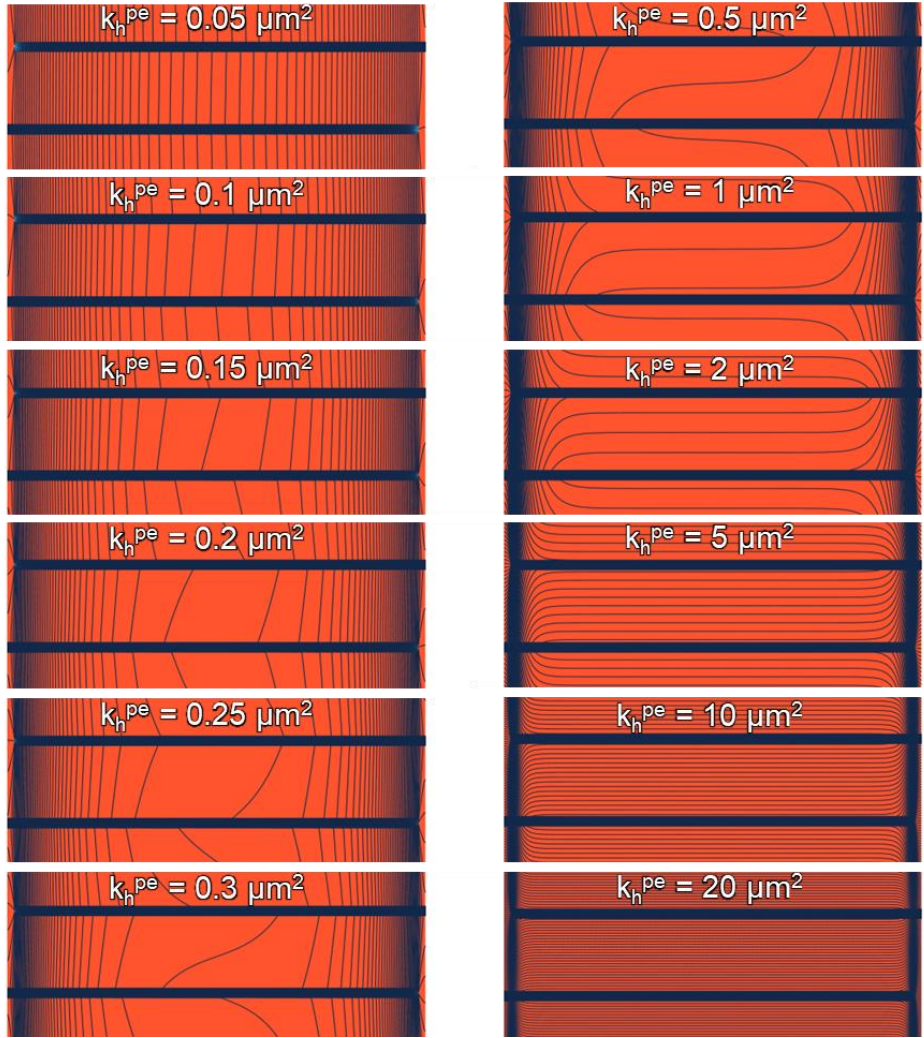

**Figure S4**: Streamlines of different interdigitated flow fields (eµ-IDFFs) with the same inter-channel distance ($s = 530$ µm), channel width ($w = 70$ µm), and gap between channel end and electrode edge ($g = 750$ µm). The electrode porosity in all cases was 45% but the porous electrode permeability $k_h^{pe}$ was varied as shown.

## Quasi-1D Analytical Model of IDFF Velocity and Permeability
### *Transverse Velocity Distribution*

We now derive the transverse velocity distribution produced within a porous electrode containing an embedded, interdigitated flow field. In contrast with our numerical modeling described in the main text, the present simplified model that neglects velocity along channels within porous electrode material enables us to identify a key non-dimensional parameter that governs the transition among IDFF designs between a transverse flow-through velocity field to a parallel velocity field.

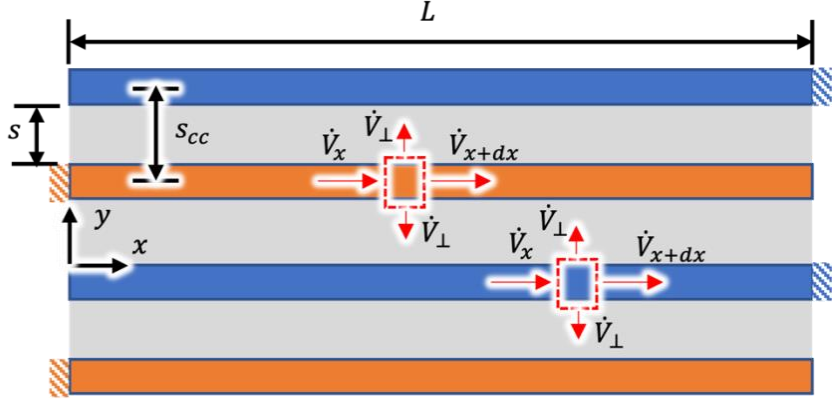

**Figure S5**: Schematic of an idealized interdigitated network of channels embedded within a porous electrode.

To derive this result, we apply conservation of fluid volume to a control volume of differential length $dx$ on both high-pressure (blue) and low-pressure (orange) channels (Fig. S5) to arrive at the following:

$$\dot{V}_x - \dot{V}_{x+dx} - 2\dot{V}_\perp = 0$$

Using Taylor series expansion of volumetric flow along a given channel, we express an ordinary differential equation that couples the cross-section averaged superficial velocity along a given channel $\langle u_{s,\parallel}\rangle$ with the thickness-averaged superficial velocity transverse to the same channel $\langle u_{s,\perp}\rangle$:

$$A_c \frac{d\langle u_{s,\parallel}\rangle}{dx} + 2\langle u_{s,\perp}\rangle h = 0$$

Here, the cross-sectional area is $A_c$ and thickness is $h$. We employ a Darcy-Darcy formulation (as already described) to relate these respective velocities to the associated pressure field. The superficial velocity along each channel depends on the channel's permeability $k_h^{ch}$ and fluid viscosity $\mu$:

$$\langle u_{s,\parallel}\rangle = -\frac{k_h^{ch}}{\mu}\frac{dp}{dx}$$

We model the transverse superficial velocity by neglecting $x$-direction contributions, in which case Darcy's law predicts linear variation of pressure with the $y$ position coordinate:

$$|\langle u_{s,\perp}\rangle| = \frac{k_h^{pe}}{\mu}\frac{(p_H - p_L)}{s}$$

Here, $k_h^{pe}$ is the hydraulic permeability of unpatterned porous electrode material and $s$ is inter-channel distance that defines the span of porous electrode material between channels. Substitution of $\langle u_{s,\parallel}\rangle$ and $\langle u_{s,\perp}\rangle$ into the above volume conservation equation produces two ordinary differential equations (ODEs) that jointly govern pressure in high- and low-pressure channels.

$$\frac{k_h^{ch} A_c}{\mu}\frac{d^2 p_H}{dx^2} - \frac{2 k_h^{pe} h}{\mu s}(p_H - p_L) = 0$$

$$\frac{k_h^{ch} A_c}{\mu}\frac{d^2 p_L}{dx^2} - \frac{2 k_h^{pe} h}{\mu s}(p_L - p_H) = 0$$

We obtain a single ODE governing the position-dependent pressure difference between adjacent channels, $\theta = p_H - p_L$, by subtracting the above equations from each other:

$$\frac{d^2\theta}{dx^2} - \frac{4 k_h^{pe} h}{k_h^{ch} s A_c}\theta = 0$$

The general solution to this equation is $\theta = c_1 e^{-mx} + c_2 e^{-mx}$ with $m = \left(4 k_h^{pe} h / k_h^{ch} s A_c\right)^{0.5}$. While in experimental practice pressure is imposed on the left end of the high-pressure channel and on the right end of the low-pressure channel, the nature of the present governing equation demands a boundary condition for the difference in pressure between these channels. We therefore invoke a symmetry boundary condition for the pressure difference, such that $\theta(x=0) = \theta(x=L) = \theta_0$. From this condition we obtain the variation of this pressure difference with position:

$$\theta(x) = \frac{\theta_0\left(\sinh(m(L-x)) + \sinh(mx)\right)}{\sinh(mL)}$$

This solution is then used directly to obtain the transverse superficial velocity distribution:

$$\langle u_{s,\perp}\rangle = \frac{k_h^{pe}\theta(x)}{\mu s} = \frac{\theta_0 k_h^{pe}\left(\sinh(m(L-x)) + \sinh(mx)\right)}{\mu s \sinh(mL)}$$

This solution lends itself naturally to dimensionless form:

$$\frac{\langle u_{s,\perp}\rangle \mu s}{\theta_0 k_h^{pe}} = u_\perp^* = \frac{\left(\sinh(m^*(1 - x/L)) + \sinh(m^* x/L)\right)}{\sinh(m^*)}$$

where $m^* = mL = \left(4 k_h^{pe} h L^2 / k_h^{ch} s A_c\right)^{0.5}$.

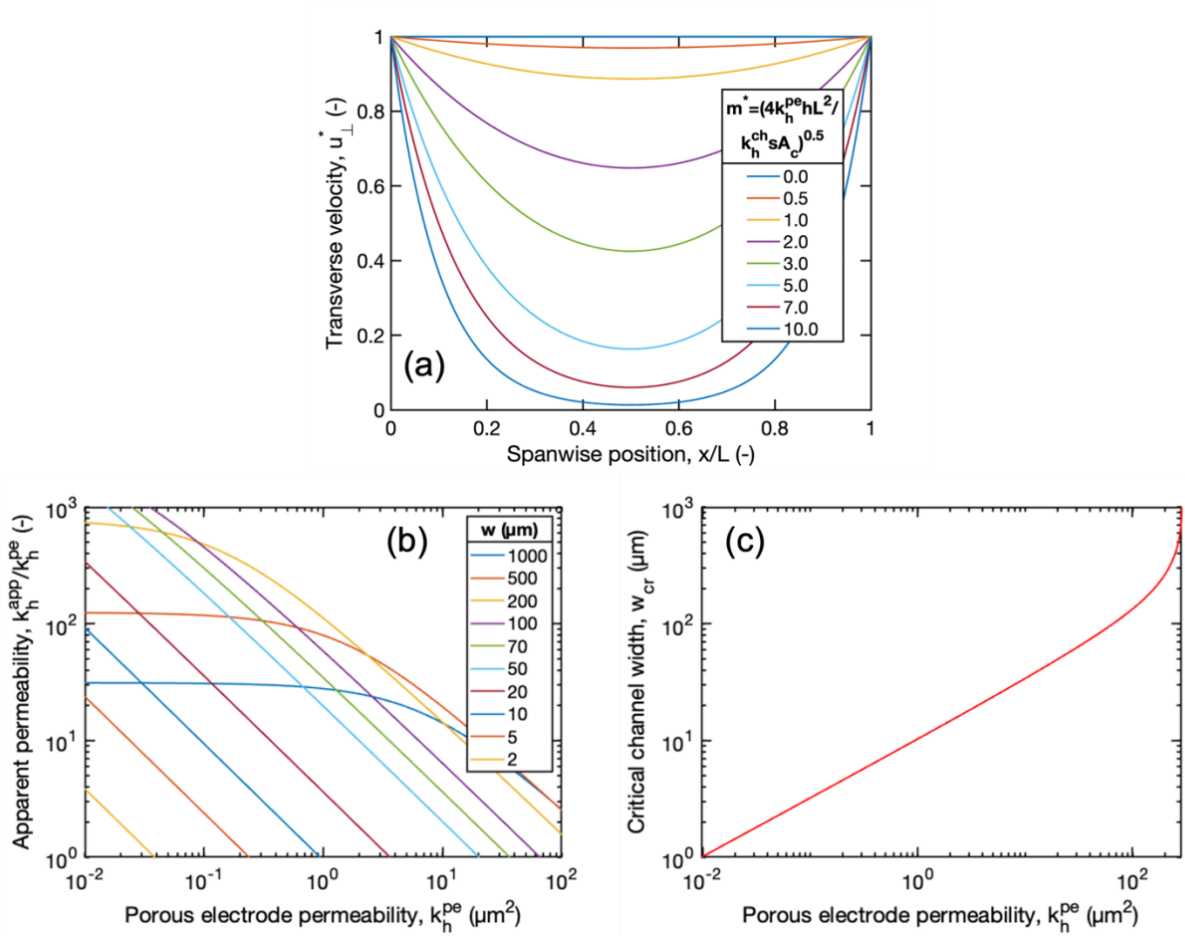

**Figure S6**: Predictions of the quasi-1D, analytical IDFF model. (a) Non-dimensional transverse velocity as a function of non-dimensional spanwise position for different $m^*$ values. (b) Normalized apparent permeability versus porous-electrode permeability with 200 μm electrode thickness, 4.5 cm electrode length, and a width to inter-channel distance ratio of 13.2%=70μm/530μm. (c) Critical channel width as a function of porous electrode permeability using the dimensional constraints.

Examination of transverse velocity profiles in Fig. S6 reveals that $m^* \ll 1$ produces uniform flow through the porous electrode, whereas $m^* \geq 10$ produces a dead zone in the porous electrode at the center of its span. To derive a condition for the rational design of such channels, we require $m^* \leq 7$, which assures that transverse velocity in the center of the porous electrode region is at least 5% of its value at the electrode's ends. For simplicity, we consider a channel cross-section that extends through the electrode's entire thickness $h$ and for which the channel's width $w$ is substantially smaller than the electrode's thickness ($w << h$), in which case $A_c = wh$ and $k_h^{ch} = w^2/12$. These assumptions produce $m^* = mL = \left(48 k_h^{pe} L^2 / w^3 s\right)^{0.5}$. Alternatively, we define a dimensionless parameter $\Xi = k_h^{pe} L^2 / (w^3 s)$,[*] which is the ratio of the characteristic longitudinal hydraulic resistance within channels to the characteristic transverse hydraulic resistance through porous electrode material. $m^*$ depends on $\Xi$ in the following way: $m^* = (48\Xi)^{0.5}$. Thus, the IDFF design criterion ($m^* \leq 7$) dictates the following condition for $\Xi$ to assure flow-through functionality of the IDFF of interest:

$$\Xi = k_h^{pe} L^2 / (w^3 s) \leq 49/48 \approx 1$$

As a result, we observe that this condition is more easily satisfied either by the electrode having low permeability, by increasing channel width, or by increasing inter-channel distance to reduce the magnitude of $\Xi$. In turn, one could also decrease channel length by decreasing electrode size to reduce $\Xi$, but doing so is expected to be impractical in many contexts.

*Apparent Permeability*

To derive an expression for the apparent permeability of the resulting IDFF we integrate the obtained transverse velocity to yield the total rate entering the high-pressure channel $\dot{V}_{ch}$:

$$\dot{V}_{ch} = 2h \int_0^L \langle u_{s,\perp} \rangle dx = \frac{2h k_h^{pe} \theta_0}{\mu s \sinh(mL)} \int_0^L \sinh(m(L-x)) + \sinh(mx) dx$$
$$= \frac{2h k_h^{pe} \theta_0}{\mu s m \sinh(mL)} [\cosh(mx) - \cosh(m(L-x))]_0^L = \frac{4h k_h^{pe} \theta_0 (\cosh(mL) - 1)}{\mu s m \sinh(mL)}$$

This equation can be used to express $\theta_0$ in terms of $\dot{V}_{ch}$:

$$\theta_0 = \mu s m \sinh(mL) \dot{V}_{ch} / \left[4h k_h^{pe} (\cosh(mL) - 1)\right]$$

We then substitute the solution for $\theta(x)$ into the volume conservation equation for the high-pressure channel to obtain an integral equation for its pressure gradient:

---

[*] We use the symbol $\Xi$ to denote this parameter because of its intuitive appeal resulting from its three parallel lines representing a high- and low-pressure channel separated by intervening porous electrode material.

$$\frac{dp_H}{dx} = \frac{2k_h^{pe} h\theta_0}{sk_h^{ch} A_c \sinh(mL)} \int \left(\sinh(m(L-x)) + \sinh(mx)\right) dx$$

$$\frac{dp_H}{dx} = \frac{2k_h^{pe} h\theta_0}{sk_h^{ch} A_c m \sinh(mL)} \left(\cosh(mx) - \cosh(m(L-x)) + \kappa\right)$$

Employing the no-flux boundary condition at the tip of the high-pressure channel subject to the Darcy's law $(dp_H/dx|_{x=L} = 0)$, we find the constant of integration as $\kappa = 1 - \cosh(mL)$. Definite integration of the above pressure gradient yields the difference in pressure between the inlet and the tip of the high-pressure channel:

$$p_H(x=0) - p_H(x=L) = \frac{2k_h^{pe} h\theta_0 \left(mL(\cosh(mL) - 1)\right)}{sk_h^{ch} A_c m^2 \sinh(mL)}$$

The total pressure difference $\Delta p = p_H(x=0) - p_H(x=L) + \theta_0$ reduces to the following after substituting $\theta_0$ in terms of $\dot{V}_{ch}$:

$$\Delta p = \left(\frac{2k_h^{pe} h \left(mL(\cosh(mL) - 1)\right)}{sk_h^{ch} A_c m^2 \sinh(mL)} + 1\right) \mu sm \sinh(mL) \dot{V}_{ch} / \left[4hk_h^{pe}(\cosh(mL) - 1)\right]$$

The apparent permeability of the interdigitated network is thereby expressed as:

$$k_h^{app} \equiv \frac{\dot{V}_{ch}\mu}{2hs_{cc}} \frac{L}{\Delta p} = \frac{4hk_h^{pe} L(\cosh(mL) - 1)}{2\left(\frac{2k_h^{pe} h \left(mL(\cosh(mL) - 1)\right)}{sk_h^{ch} A_c m^2 \sinh(mL)} + 1\right) mhs_{cc}s \sinh(mL)}$$

Here, $s_{cc} = s + w$ is the center-to-center span between channels. Further simplification using the definition of $m$ yields the following expression for the normalized apparent permeability $k_h^{app}/k_h^{pe}$, as predicted by this quasi-1D model:

$$\frac{k_h^{app}}{k_h^{pe}} = k_{h,max}^{app,*} \frac{4(\cosh(m^*) - 1)}{(m^*)^2(\cosh(m^*) - 1) + 2(m^*)\sinh(m^*)}$$

$$= k_{h,max}^{app,*} \frac{4(\coth(m^*) - \operatorname{csch}(m^*))}{(m^*)^2(\coth(m^*) - \operatorname{csch}(m^*)) + 2m^*}$$

Here, $k_{h,max}^{app,*} = L^2/[s(s+w)]$ is the upper bound of normalized apparent permeability derived in the limit of vanishing hydraulic resistance within channels when $L = L_{ch} \approx L_e$, as described in our asymptotic scaling analysis that precedes this section.

## Laser-Patterning of Intercalation Electrodes
### *Engraving Procedures*

As stated in the main text, a channel width of 70 μm, a spacing between two channels of 530 μm, and a channel depth of 200 μm was chosen for our eμ-IDFFs. Hence, over the electrode area of 4.5×4.5 cm consists of 74 microchannels, as shown in Fig. S7. To engrave this eµ-IDFF into our electrodes, a Trotec Speedy 400 Flexx laser was used. This instrument is a dual head laser engraver with a $CO_2$ laser and a 50W Yb- doped fiber laser, but only the fiber laser head was used.

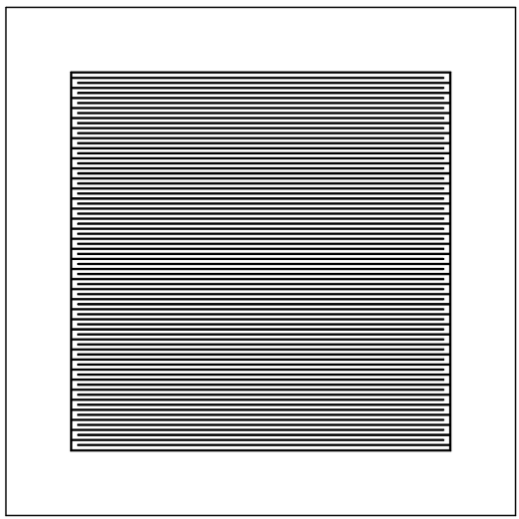

**Figure S7**: A sketch of the full eµ-IDFF that was engraved into electrodes. The dimension across the eµ-IDFF is 4.5 cm.

Channels were engraved by moving the laser along a series of parallel passes. For all laser passes, 20% of the maximum laser power (50W) was used with the speed of the laser being 1 m s$^{-1}$ and the repetition rate being 100 kHz. These settings result in a fluence of 0.895 J cm$^{-2}$ (± 0.158 J cm$^{-2}$), which was calculated as $P\tau/A_{spot}$, where $P$ is the laser power, $A_{spot}$ is the area of the laser spot, and $\tau$ is the laser pulse width. $A_{spot}$ was estimated using averaged diameter of holes that were ablated in an 8 µm thick sheet of aluminum foil (All-Foils, Inc.) (Fig. S8). These discrete holes were obtained by increasing the laser power and the scanning speed to space out the laser pulse train. The average hole diameter was determined to be 11.9 ± 1.2 µm. Since the specific pulse width of the Trotec laser could not be easily found, we assumed the pulse width to be 100 ns, which falls within the range reported in the manual (1-1000 ns).

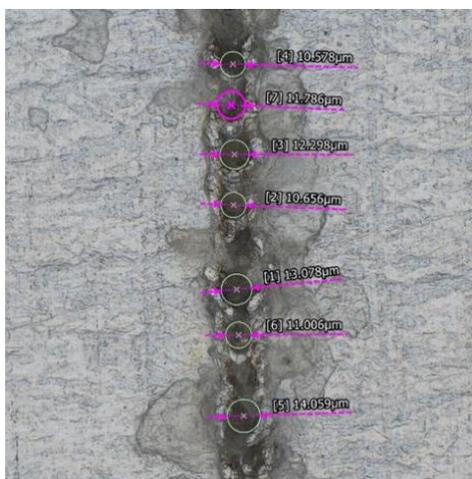

**Figure S8**: Laser ablation of holes in 8 µm thick aluminum with measured hole diameters.

To create a single channel with sufficient width and depth, three adjacent laser passes were performed on the electrode of interest with a spacing of 25 µm between the path that defined each pass, as depicted in Fig. S9. These three adjacent laser paths were repeated twice to create a given channel.

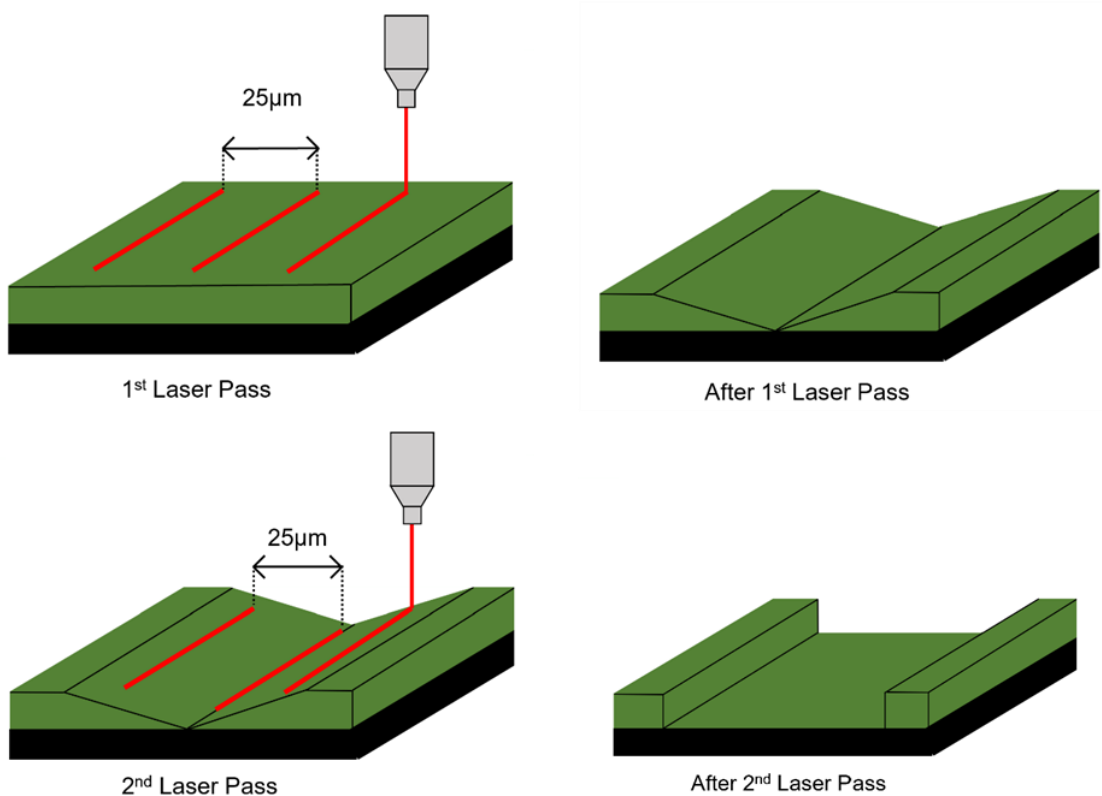

**Figure S9**: Idealized schematic for the engraving of a single microchannel. With the above mentioned fluence, the laser ablates the electrode composite (green) but does not cut through the graphite (black).

*Engraving of Water-Imbibed Electrodes for Improved Channel Quality*

     Prior to engraving, the pores of all samples were imbibed with deionized (DI) water to improve channel resolution and smoothness. As shown in Fig. S10, the channels produced on water-imbibed electrodes were consistently smaller than those produced on dry electrodes at the same laser power, regardless of whether they were made with three laser passes or one. In addition, the root-mean-square roughness along the centerline and side walls of the channels were lower when the water imbibed approach was used (Fig. S11).

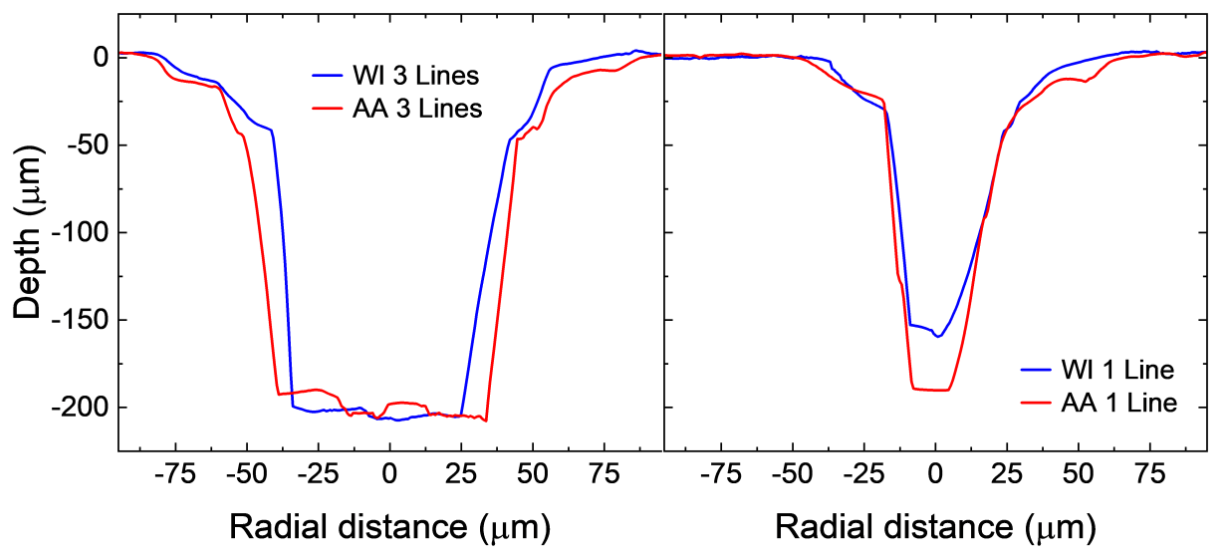

**Figure S10**: Example of cross-sections of channels produced on water-imbibed (WI) electrodes and dry electrodes with air assisted (AA) using three laser paths (left) and one laser path (right).

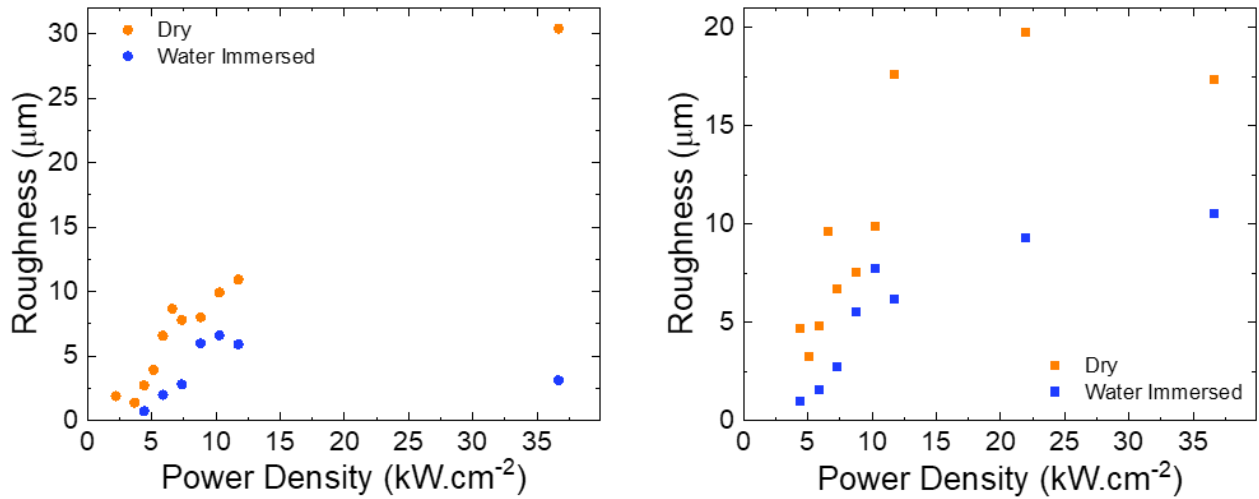

**Figure S11**: Root-mean-square roughness along the centerline (left) and sidewalls (right) of channels created with water-imbibed and dry electrodes at various laser power densities.

*eµ-IDFF Characterization*

The geometry and microstructure of channels were characterized using a laser profilometer (Keyence VK-X1000 3D Laser Scanning Confocal Microscope). As mentioned in the main text, three pairs of electrodes with different PBA loading levels (15 mg cm$^{-2}$, 19 mg cm$^{-2}$, and 21 mg cm$^{-2}$) were fabricated and patterned with the chosen eµ-IDFF design. Profilometry was performed on one electrode of each pair to characterize the resulting eµ-IDFF. The ideal channel shape was a rectangular channel with a width and depth of 70 µm and 200 µm, respectively, but the actual channel shape was trapezoidal (Fig. S10) due to the laser beam having a certain divergence angle and a certain intensity distribution. Hence, the cross-sectional area of the channel was a more effective classification parameter, since similar cross-sectional area would lead to similar pressure drop according to Darcy's law.

Three images were taken of each electrode, including two images of the channels near the corners of each electrode and one image of a channel at the center of each electrode. Even with consistent laser processing settings, we observed variability of these dimensions across a given electrode. The channels in the corners of the electrodes had consistently higher cross-sectional areas than the center. Table S5 shows the cross-sectional areas for the electrodes.

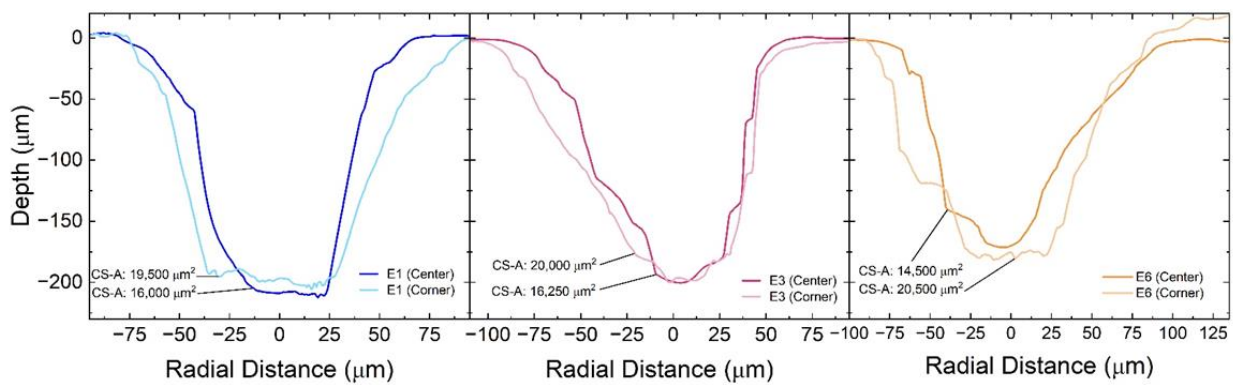

**Figure S12**: Channel profiles near the edge and in the center of electrodes with PBA loading of 15 mg cm$^{-2}$ (left), 19 mg cm$^{-2}$ (middle), and 21 mg cm$^{-2}$ (right). The cross-section area (CS-A) for each profile are shown in the plot.

Fig. S12 shows how the channels at the corners of the electrodes have a systematically larger cross-sectional area than at the center. The x and y directions of the Trotec Speedy 400 Flexx laser are controlled by a mechanical gantry system. At the beginning and end of the cutting paths the gantry system will have some amount of mechanical inertia. Since the laser fires during the entire laser path, the laser will spend more time at the end of the channels as it accelerates and decelerates due to this inertia. This effect likely resulted in a larger cross-sectional area at the edges and corners of the electrodes.

**Table S1**: Statistics of cross-sectional areas of channels of three engraved electrodes

| Electrode sample | Cross-section area (corner) (µm$^2$) | | | Cross-section area (center) (µm$^2$) | | |
|---|---|---|---|---|---|---|
| | Max | Min | Med | Max | Min | Med |
| L1 | 21000 | 18000 | 19500 | 17000 | 15000 | 16000 |
| M3 | 22000 | 18000 | 20000 | 17000 | 15500 | 16250 |
| H6 | 22000 | 19000 | 20500 | 15500 | 13500 | 14500 |

## Design and Layout of the FDI Flow Cell Used

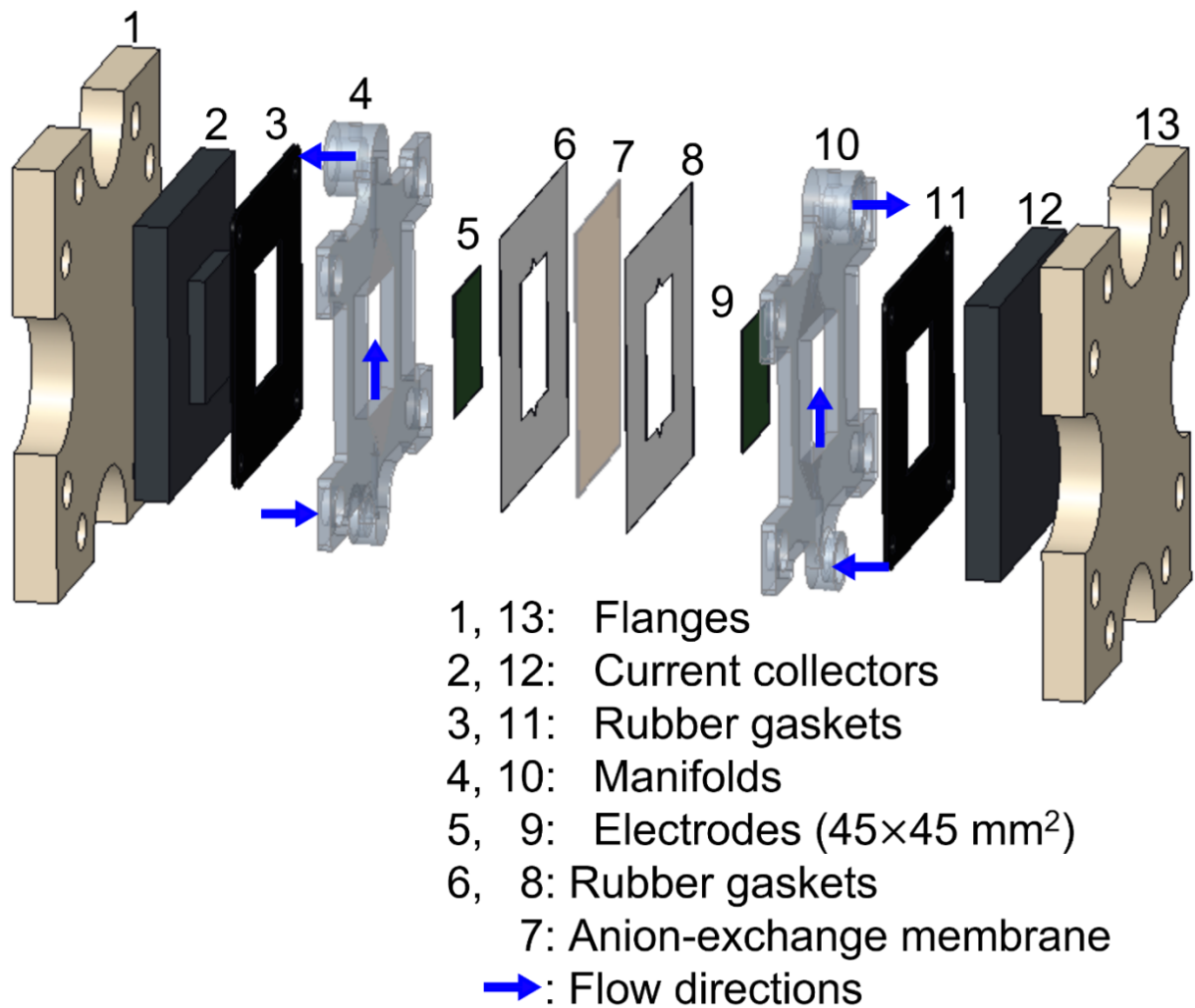

1, 13: Flanges
2, 12: Current collectors
3, 11: Rubber gaskets
4, 10: Manifolds
5, 9: Electrodes (45×45 mm$^2$)
6, 8: Rubber gaskets
7: Anion-exchange membrane
→: Flow directions

**Figure S13**: Components of the FDI flow cell.

# Details of Desalination Performance

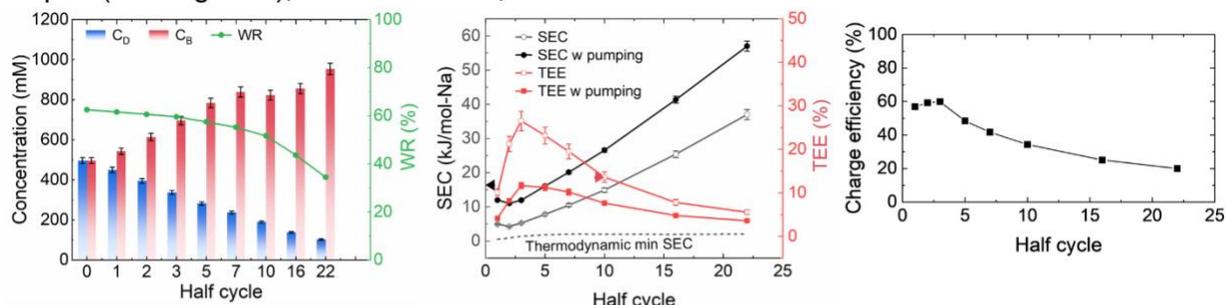

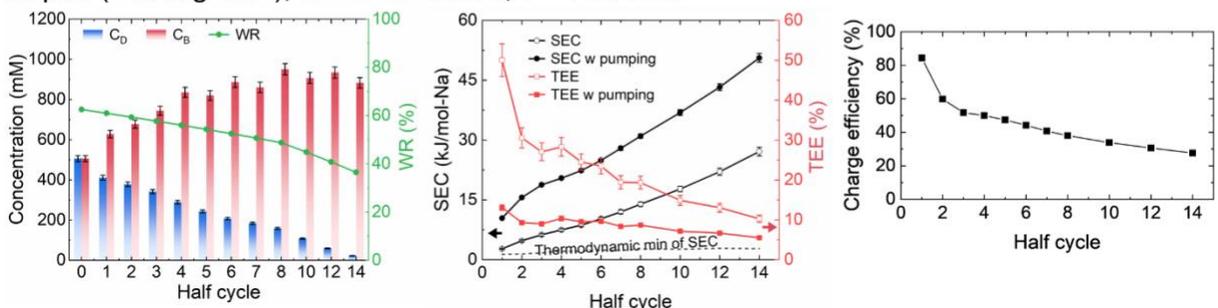

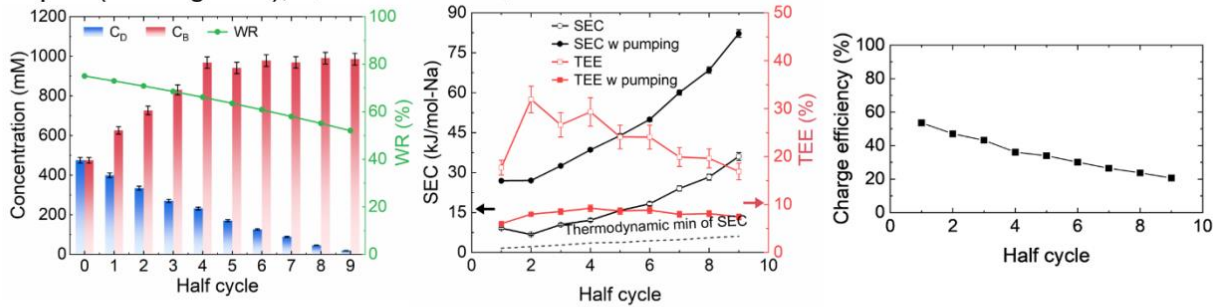

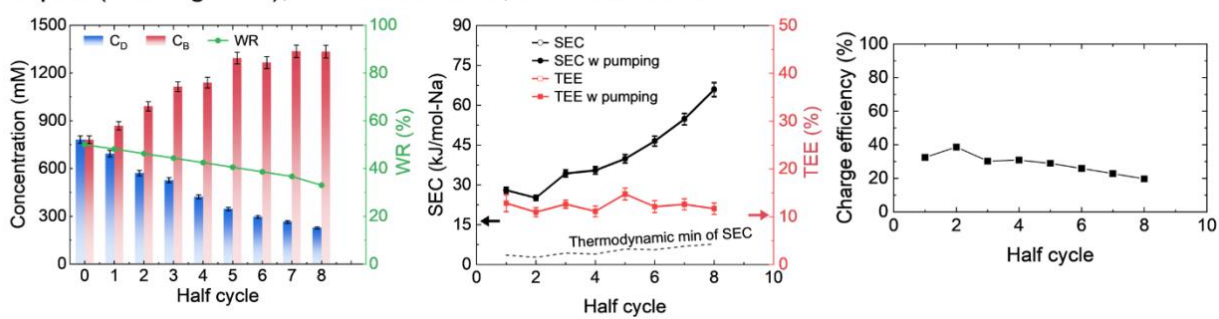

**Figure S14**: Detailed data for desalination experiments in Fig. 5a of the main text. Salt concentrations in diluate and brine reservoirs, specific energy consumption (SEC), thermodynamic energy efficiency (TEE), and charge efficiency versus half-cycle in desalination experiments with the electrodes and conditions shown.

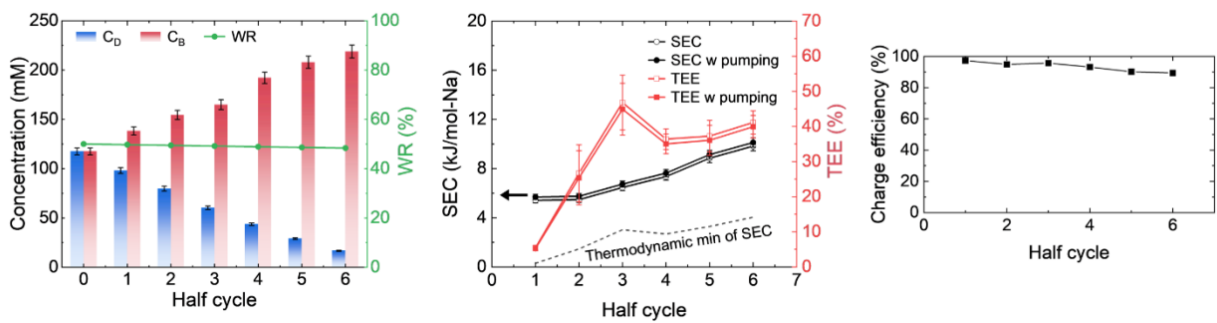

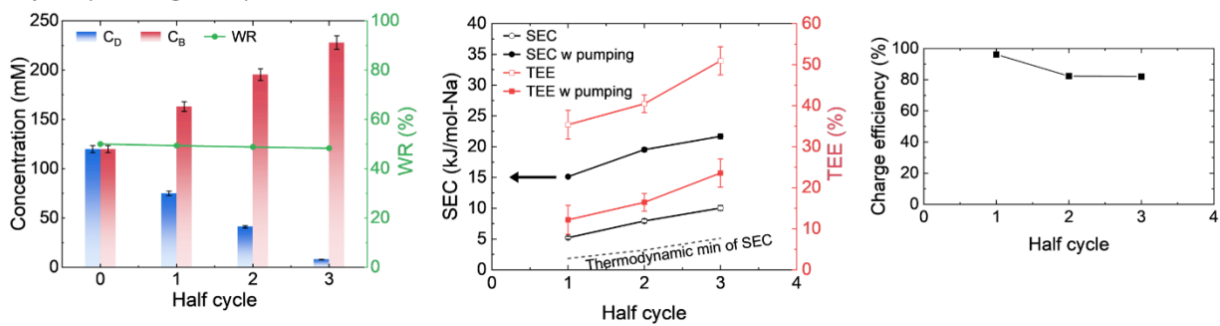

**Figure S15**: Detailed data for desalination experiments in Fig. 5b of the main text. Salt concentrations in diluate and brine reservoirs, SEC, TEE, and charge efficiency versus half-cycle in desalination experiments with the electrodes and conditions shown.

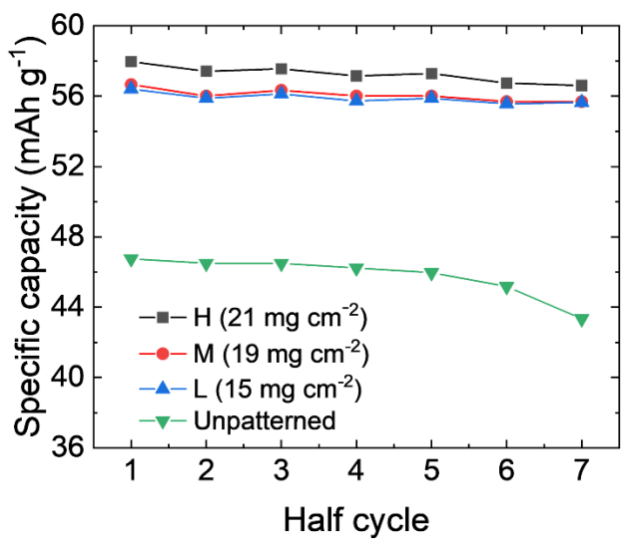

**Figure S16**. Specific capacity of patterned and unpatterned electrodes in an FDI cell with 500 mM NaCl feed concentration.

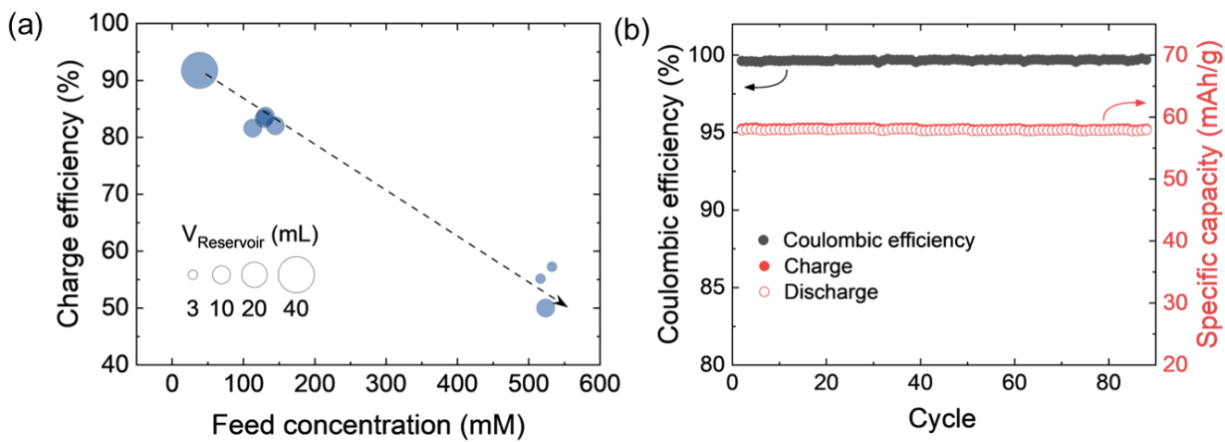

**Figure S17**. (a) Dependence of charge efficiency on feed concentration and reservoir volume. (b) Coulombic efficiency and specific capacity of electrodes. These experiments were performed using H-pair electrodes with 21 mg cm$^{-2}$ areal loading.

Prior to each desalination experiment we assembled the FDI flow cell with a fresh piece of as-received Neosepta AMX anion exchange membrane (AEM) material, and each AEM was inspected visually after each desalination experiment *ex situ*. The AEMs used for each desalination experiment used to obtain the data in Fig. 4 showed no changes in color or texture. However, weak yellow discoloration of an AEM (Fig. S18) was observed after H-pair electrodes were cycled for 90 cycles in 500 mM NaCl to obtain the data in Fig. S7b. Previous work[3] implicated changes of Neosepta AMX from transparent to translucent yellow, orange, red, or black with dehydrochlorination of poly(vinyl chloride) to polyene due to alkali attack. We hypothesize that side reactions that produce OH$^-$ during the instants that cell potential approaches its extreme (±0.4V) could be responsible for this degradative effect. Because the electrodes and the AEM are compressed within the flow cell, solution near the electrode/AEM interface may have reduced local fluid permeability, limiting access of fresh salt solution to their interface, leading to the local depletion of salt and thus promoting side reactions instead of cation intercalation. These observations therefore motivate detailed investigation of membrane longevity in FDI and the mechanisms that determine it. Because longer exposure time and higher NaOH concentration were shown previously to result in darker hues of Neosepta AMX undergoing alkali attack, the light yellow coloration observed in our present experiments indicates minimal AEM degradation that is likely a result of the galvanostatic cycling conditions (constant current) that we used in this work, rather than potentiostatic conditions (constant potential) that would hold the cell at extreme potentials for much longer time periods.

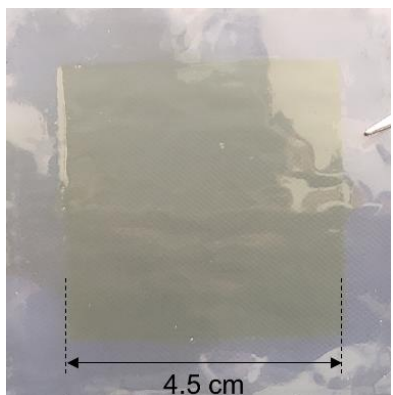

4.5 cm

**Figure S18**. Photo of Neosepta AMX AEM after 90 cycles in a flow cell where the AEM was sandwiched between H-pair electrodes, as described in the caption of Fig. S7b.

**Table S2.** Data for Fig. 1a in the main text.

| Author, year | Flow configuration | Electrode capacity (mAh) | Salt removal (mM) | Feed concentration (mM) | Ref.* |
|---|---|---|---|---|---|
| Kim et al., 2017 | Flow by | 0.24745 | 7.5 | 25 | 31 |
| Ahn et al., 2021 | Flow by | 5.04 | 100 | 500 | 32 |
| Son et al., 2020 | flow-through vs flow-by | 1.68 | 15 | 50 | 30 |
| Pothanamkandathil et al., 2020 | flow by | 0.83333 | 5 | 20 | 29 |
| Porada, 2017 | flow by | 47 | 0.5 | 20 | 33 |
| Lee et al., 2017 | batch | 5.058 | 190.8 | 477 | 17 |
| Reale 2019 | flow through | 0.6 | 27 | 100 | 28 |
| Reale 2021 | flow through | 1.84 | 104 | 200 | 26 |
| Reale 2021 | flow through | 1.84 | 90 | 100 | 26 |
| This work | | 21.5 | 455.5 | 474.6 | |
| This work | | 21.5 | 112 | 119.8 | |
| This work | | 21.5 | 553 | 780 | |

* The reference number here refers to that in the main text.

**Table S3**: Data for Fig. 1b in the main text.

| Flow Field Type | Notes | Cell Area (cm$^2$) | Channel Width (mm) | Flow Path (mm) | Ref* |
|---|---|---|---|---|---|
| SFF | - | 25 | 1 | 1 | 50 |
| PFF & IDFF | - | 23 | 1.1 | 0.89 | 58 |
| SFF & IDFF | - | N/A | 1 | 1 | 60 |
| SFF & IDFF | - | 100-625 | 1-4 | 1-8 | 62 |
| MSFF | Modified Serpentine | N/A | 2 | 2 | 46 |
| MSFF | Modified Serpentine | 16-100 | 4 | 4 | 45 |
| MSFF | Modified Serpentine | 2200 | 5 | 2 | 66 |
| IDFF | - | 25 | 1 | 0.8 | 51 |
| IDFF & MIDFF | Modified Interdigitated | 900-1500 | 1 | 2-5 | 53 |
| SFF | - | 1500 | 3-5 | 2-3 | 65 |
| SFF & IDFF | - | 0.23 | 0.8 | 0.8 | 57 |
| IDFF | - | 10.24 | 1 | 1 | 56 |
| IDFF | - | N/A | 3 | 1-2.3 | 49 |
| IDFF | - | 4 | 1 | 1 | 59 |
| IDFF | - | N/A | 1 | 1 | 47 |
| SFF | - | 400-900 | 3-5 | 2-3 | 64 |
| PFF & IDFF | - | 10 | 1 | 1 | 52 |
| IDFF & MIDFF | Modified Interdigitated | 9-900 | 1 | 1 | 68 |
| IDFF & MIDFF | Modified Interdigitated | 39.84 | 1-3 | 1 | 44 |
| SFF & IDFF | - | N/A | 1 | 1 | 63 |
| IDFF | - | 625 | 2 | 3 | 61 |
| IDFF | - | N/A | 1 | 1 | 48 |
| SFF | - | N/A | 1 | 1 | 55 |
| MPFF | Modified Parallel | N/A | 0.2-0.8 | 0.25 | 70 |
| SFF & IDFF | - | 4 | 1 | 1 | 54 |
| PFF | - | 32 | 2 | 1 | 69 |
| IDFF | - | 20.25 | 0.07 | 0.5 | This work |

* The reference number here refers to that in the main text.

**Table S4**: Data for Fig. 1c in the main text.

| | CDI (min) | CDI (max) | FDI flow-by (min) | FDI flow-by (max) | FDI flow-through (min) | FDI flow-through (max) | This work (min) | This work (max) |
|---|---|---|---|---|---|---|---|---|
| TEE (%) | 0.1 | 3 | 5 | 40 | 5 | 40 | 3 | 40 |
| $E_{pump}$ (W h m$^{-3}$) | 0 | 0 | 0 | 0 | 0 | 50 | 0.005 | 1.2 |
| SAC (mg g$^{-1}$) | 10 | 20 | 40 | 100 | 40 | 100 | 132.8 | 161.4 |
| SAR (mg g$^{-1}$min$^{-1}$) | 0.1 | 1 | 0.01 | 0.02 | 0.01 | 0.02 | 0.22 | 1.16 |
| SR (mM) | 0 | 5 | 0.5 | 100 | 15 | 104 | 394 | 554 |

**Table S5:** Raw data for Figs. 5a and 5b in the main text.

| Electrodes | Feed concentration (mM) | Current density (mA cm$^{-2}$) | Reservoir volume (mL) | Flow rate (mL min$^{-1}$) | Dilutate concentraiton (mM) | Water transport (L m$^{-2}$h$^{-1}$) |
|---|---|---|---|---|---|---|
| L | 496.4 | 1 | 5 | 5 | 102.4 | 0.095 |
| M | 496.4 | 1 | 5 | 5 | 23.7 | 0.107 |
| H | 496.4 | 1 | 5 | 5 | 19.1 | 0.148 |
| H | 780.9 | 8.5 | 5 | 1 | 227.5 | 0.159 |
| M | 118.0 | 1 | 30 | 5 | 16.7 | - |
| H | 119.8 | 1 | 30 | 5 | 7.8 | - |